\documentclass[11pt,a4paper]{article}

\usepackage{graphicx}
\usepackage{amssymb}
\usepackage{textcomp}
\usepackage{amsmath}
\usepackage{geometry}                
\usepackage{graphicx}
\usepackage{amssymb}
\usepackage{epstopdf}
\usepackage{geometry}                
\usepackage{graphicx}
\usepackage{amssymb}
\usepackage{epstopdf}
\usepackage{bm}
\usepackage{times}
\usepackage{epsfig}
\usepackage{color}
\usepackage{array,multirow}
\usepackage{jheppub}

\newcommand{\pmns}{{\mbox{\tiny PMNS}}}

\newcommand{\trix}[1]{\left(\begin{array}{#1}}
\newcommand{\notrix}{\end{array}\right)}
\newcommand{\comment}[1]{}

\def\beq{\begin{equation}}
\def\eeq{\end{equation}}
\def\bea{\begin{eqnarray}}
\def\eea{\end{eqnarray}}

\usepackage[normalem]{ulem}

\newcommand{\madgraph}{\textsc{MG5\_aMC@NLO}}
\usepackage[T1]{fontenc}

\usepackage[utf8]{inputenc}

\usepackage{mathtools} 

\usepackage{amsfonts}

\usepackage{ amssymb }

\usepackage[toc,page,title]{appendix}

\usepackage{xcolor}

\usepackage[compat=1.1.0]{tikz-feynman}

\usepackage{graphicx}

\usepackage{subcaption}

\usepackage{float}

\usepackage{caption}

\usepackage{breqn}

\usepackage{amsmath}

\usepackage{etoolbox}
\makeatletter
\patchcmd{\maketitle}{\@fpheader}{}{}{}
\makeatother

\usepackage[compat=1.1.0]{tikz-feynman}
\numberwithin{equation}{section}
\title{\Large  {{\bf{$R$-parity Violating Decays of\\ 
 Bino Neutralino\\
LSPs at the LHC}}}}
\author{{Sebastian Dumitru, Christian Herwig and Burt A.~Ovrut} \\[2mm
]
    {\it Department of Physics and Astronomy, University of Pennsylvania} \\
   {\it Philadelphia, PA 19104--6396}\\
   [4mm]
}
\date{\today}

{\footnotetext{\mbox{sdumitru@sas.upenn.edu, ~herwig@sas.upenn.edu ~ovrut@elcapitan.hep.upenn.edu} }}
\abstract{
The $R$-parity violating decays of Bino neutralino LSPs are analyzed within the context of the $B-L$ MSSM ``heterotic standard model''. These LSPs correspond to statistically determined initial soft supersymmetry breaking parameters which, when evolved using the renormalization group equations, lead to an effective theory satisfying all phenomenological requirements; including the observed electroweak vector boson masses and the Higgs mass. The explicit RPV decay channels of these LSPs into standard model particles, the analytic and numerical decay rates and the associated branching ratios are presented. The analysis of these quantities breaks into two separate calculations; first, for Bino neutralino LSPs with mass larger than $M_{W^{\pm}}$ and, second, when the Bino neutralino mass is smaller than the electroweak scale. The RPV decay processes in both of these regions is analyzed in detail. The decay lengths of these RPV interactions are discussed. It is shown that for heavy Bino neutralino LSPs the vast majority of these decays are ``prompt'', although a small, but calculable, number correspond to ``displaced'' decays of various lengths. The situation is reversed for light Bino LSPs, only a small number of which can RPV decay promptly. The relation of these results to the neutrino hierarchy--either normal or inverted--is discussed in detail.  }

\begin{document}

\maketitle

\newpage

\newpage
%
\section{Introduction}

In a series of papers \cite{Braun:2005nv,Ambroso:2009jd,Ambroso:2009sc,Ambroso:2010pe,Ovrut:2012wg,Braun:2013wr,Ovrut:2014rba,Ovrut:2015uea}, a minimal extension of the $R$-parity invariant MSSM was presented. This extended theory, in addition to the usual structure of the MSSM, introduced three right-handed neutrino chiral multiplets--one per family--and gauged the global $B-L$ symmetry, thus introducing an anomaly free $B-L$ vector supermultiplet. This $B-L$ MSSM was shown to arise ``from the top down'' as the observable sector of a supersymmetric compactification  to four-dimensions \cite{Evans:1985vb,Molera:1989cp,Donagi:1999jp,Donagi:2000fw,Donagi:2004ia,Braun:2005zv,Anderson:2009nt} of $E_{8} \times E_{8}$ heterotic M-theory \cite{Horava:1996ma,Lukas:1997fg,Lukas:1998ew,Lukas:1998yy,Lukas:1998tt}. It was also introduced ``from the bottom up'' point of view as an effective field theory extension of the MSSM \cite{FileviezPerez:2008sx,Barger:2008wn,FP:2009gr,Everett:2009vy,FileviezPerez:2012mj,Perez:2013kla}. When the $N=1$ supersymmetry is spontaneously broken, this $B-L$ MSSM--for a wide range of soft supersymmetry breaking parameters--has been shown to satisfy all present phenomenological requirements \cite{Ovrut:2014rba,Ovrut:2015uea}. Specifically, 1) the gauged $B-L$ symmetry is radiatively broken at a sufficiently high scale to satisfy the lower bound on the $Z_{R}$ vector boson while still sufficiently suppressing both proton decay and lepton number violation, 2) electroweak symmetry is radiatively broken and gives the precise masses for the electroweak gauge bosons $W^{\pm}$ and $Z^{0}$, as well as the correct Higgs boson mass, and 3) all sparticles masses exceed their present experimental bounds. The initial soft supersymetry breaking parameters that lead to completely realistic low-energy effective theories are called ``viable'' initial points. They are determined by statistically scanning over a vast range of soft supersymmetry breaking parameters, scaling the theory down to the electroweak scale using the renormalization group, and then choosing those initial parameters that lead to completely realistic results. We conclude that the $B-L$ MSSM appears to be a well-motivated candidate for an $N=1$ supersymmetric extension of the standard model of particle physics.

As first discussed in discussed in \cite{Barger:2008wn} and then proven in detail using radiative breaking via the RGEs \cite{Ambroso:2009jd}, soft supersymmetry breaking leads to the violation of the $B-L$ symmetry by inducing a non-vanishing VEV for the third family right-handed sneutrino. Since this VEV violates $B-L$ symmetry, it spontaneously breaks $R$-parity. Although this violation is too small to cause detectable proton decay and lepton number violation, it is sufficiently large to lead to potentially observable RPV interactions in the effective theory below the $B-L$ breaking scale. A detailed discussion of this violation of $R$-parity,  its relationship to the neutrino mass hierarchy, the associated RPV parameters and their relative strength was given in \cite{Ovrut:2015uea}. For a generic sparticle, the RPV decays are sufficiently smaller than the $R$-parity conserving interactions, that the associated RPV decays are difficult to observe experimentally. However for the ``lightest supersymmetric particle'', the so-called LSP, no $R$-parity conserving decays are possible. Depending on the initial values of the soft supersymmetry breaking parameters, such theories will have different LSP candidates. When $R$-parity is spontaneously broken, as it is in the $B-L$ MSSM, the chosen LSP will decay via exactly calculable RPV interactions into standard model particles. Since there are no $R$-parity preserving interactions for an LSP, the RPV decay modes will be explicitly observable experimentally. To concretely realize these results, the LSP associated with each ``viable'' set of initial soft supersymmetry breaking parameters has been determined using the renormalization group, and the results displayed statistically by scanning over all ``viable'' parameters.  The results have been presented in \cite{Ovrut:2015uea}.

It is clear from these statistical scans that there are many different LSPs associated with the viable soft supersymmetry breaking parameters. However, it is equally clear that some LSPs are far more abundant than others; several sparticle species never occur as an LSP at all, while some others occur, but are relatively rare. The first study of the RPV decays of LSPs in the $B-L$ MSSM was carried out in \cite{Marshall:2014kea}. In this paper, the viable points were chosen so as to produce the lightest stop sparticle as the LSP. Although the relative abundance of a light stop LSP in viable models is relatively small, this mode was chosen since the cross section to produce light stop pairs is significant in proton-proton ($pp$) collisions at the Large Hadron Collider (LHC). A complete analysis of the RPV decay modes of the light stop to standard model particles, the decay rates, the associated branching ratios and the relation of these RPV decays to the neutrino mass hierarchy was carried out in \cite{Marshall:2014cwa}. These decays were searched for in the Run 1 and Run 2 data from ATLAS, with no significant excess of events observed~\cite{Aaboud:2017opj}.
Although these searches substantially improved constraints on the allowed range of stop masses, models with an LSP stop as light as 600 GeV remain consistent with the experimental data. These decay modes continue to be searched for at ATLAS, taking advantage of larger collected data sets, improved experimental techniques, and a potential increase in collision energy.

Following this study, it was observed that there are a number of different LSPs in viable $B-L$ MSSM theories that are statistically far more prevalent than the light stop. For example, various chargino and neutralino LSPs occur far more often than do stops. Therefore, in \cite{Dumitru:2018jyb}, a complete theoretical analysis of the RPV decays of chargino and neutralino LSPs was carried out. By diagonalizing the associated mass matrices, the various chargino and neutralino mass eigenstates were computed and their decay processes to standard model particles determined. The {\it analytic} formulas for their RPV decay rates and branching ratios were analyzed in detail and presented for any chosen chargino and neutralino LSP. This work presented the detailed {\it theoretical} underpinnings for any chargino and neutralino decay, but did not analyze any specific choices. The first such specific computation was carried out in \cite{Dumitru:2018nct} where, using the general formulas developed in \cite{Dumitru:2018jyb}, the decay modes, decay rates and branching ratios for Wino charginos and Wino neutralinos were computed and their experimental predictions presented. As was done in the case of the light stop LSP, the relationship of these RPV decays to the neutrino mass hierarchy was analyzed. This recent set of calculations paves the way for novel experimental searches to be conducted at the LHC.

The Wino chargino and Wino neutralino LSPs in \cite{Dumitru:2018nct} were chosen because their decay products are readily observable by ATLAS at the LHC. However, it was noticed that one particular LSP, the neutral fermionic superpartner ${\tilde{B}}$ of the hypercharge gauge field $Y$--referred to as the ``Bino''-- was a much more prevalent LSP associated with viable $B-L$ MSSM models. A statistical analysis showed that the Bino was approximately a factor of 10 more likely to occur that either a Wino chargino or a Wino neutralino. In fact, it is the most likely LSP to occur for any viable initial conditions. It seems well-motivated, therefore, to apply the results of \cite{Dumitru:2018jyb} to the Bino and to determine its decay modes to standard model particles, and the associated decay rates and branching ratios--as well as the relationship of these decays to the neutrino mass hierarchy. However, an {\it important new phenomenon} occurs for the Bino which is unique among all chargino and neutralino LSPs. That is, a careful analysis of the mass eigenstates of these LSPs shows that the Bino mass, although generically near, or above, the electroweak breaking scale, can be fine-tuned to be smaller than this scale. In fact, for sufficient fine-tuning it can be made to be arbitrarily small and even to vanish. On the other hand, this analysis reveals that the masses of all other charginos and neutralinos can never be smaller than the electroweak scale. Since the RPV decay products of the Bino must include either a $W^{\pm}$, a $Z^{0}$ or a neutral Higgs boson, the ``usual'' decays to standard model particles can no longer occur when the Bino mass drops below the electroweak scale. However, it can still decay via more complicated processes involving ``off-shell'' weak  interaction vector bosons or the Higgs. The decay products of these processes are also readily observable at the LHC. It is of interest, therefore, to also analyze these more complicated decays of a light Bino, and to compute their decay rates and branching ratios. Therefore, to summarize: the analysis of LSP Bino RPV decays carried out in this paper breaks naturally into two parts--1) the RPV decays of a Bino with mass $M_{W^{\pm}} < M_{B}$ via ``on-shell'' $W^{\pm}$, $Z^{0}$ or $h^{0}$ bosons. Such decays are similar in form to those of both Wino chargino and Wino neutralino decays studied in \cite{Dumitru:2018nct} and 2) the RPV decays of a Bino with mass $M_{B} < M_{W^{\pm}}$ via ``off-shell'' $W^{\pm}$, $Z^{0}$ or $h^{0}$ bosons. These decays are more complicated and the decay rates and branching ratios are considerably suppressed relative to the on-shell case. Be that as it may, they lead to interesting signatures, which can offer unique signals for an ATLAS search.

It is the purpose of this paper to analyze the RPV decays of a Bino LSP--for the Bino mass both above and below the electroweak scale. Specifically, we will do the following. In Section 2, following the procedure presented in \cite{Ovrut:2014rba,Ovrut:2015uea}, we analyze the low-energy predictions for each of $10^{8}$ sets of soft supersymmetry breaking parameters--each set generated randomly within a fixed mass range chosen so as to allow experimental detection of RPV decays at the LHC \cite{Dumitru:2018jyb}. A graph indicating all ``viable'' initial points--that is, those sets of initial conditions whose low energy predictions satisfy all present experimental bounds--is displayed. This data is used to construct and present a histogram of all possible LSPs associated with these viable initial points. As will become clear, the most abundant possible LSP associated with the viable points is the Bino. This arises as the LSP of a subset of the viable points, which will then also be graphically displayed. In Section 3, using the generic results presented in \cite{Dumitru:2018jyb}, we will explicitly give the eigenstate formula for the mass of the Bino in terms of the other parameters in the theory. The mass formula for the Bino will be compared to the explicit mass formula for a generic Wino chargino or Wino neutralino. It will be shown that, although a Wino chargino or a Wino neutrino mass cannot be smaller than the electroweak scale, the Bino mass can, if so desired, have an arbitrarily small mass.
In Section 4 we begin our formal analysis of the LSP Bino by presenting its explicit decay channels for a Bino with mass exceeding the electroweak scale. The RPV decay channels, decay rates, and branching ratios for these heavier Binos will be presented and analyzed for both a normal and inverted neutrino hierarchy. In Section 5, we continue the analysis of the LSP Bino RPV decay channels, decay rates and branching ratios--for both a normal and inverted neutrino hierarchy-- but now for the case when the Bino mass is less than the electroweak scale. In both Sections 4 and 5, a detailed analysis of the decay length associated with the various Bino decays, for both heavy and light Bino masses, will be given--explicitly defining and discussing so-called ``prompt'' decays and those decays with ``displaced'' vertices. We present a formal Conclusion of these analyses in Section 6.
Finally, the analysis in this paper uses the notation and formalism for the $B-L$ MSSM presented in a series of papers \cite{Ovrut:2015uea,Dumitru:2018jyb}.  The reader is referred specifically to \cite{Dumitru:2018jyb} for the precise notation and formalism particularly relevant to the definition and decay processes of the Bino LSP. For clarity, we have summarized the relevant notation in Appendix A of this paper. The analytic expressions for generic neutralino decay rates were presented in detail in \cite{Dumitru:2018jyb}. They are reproduced here in Appendix B for completeness.

Finally, we want to make three important statements concerning the computations in, and the context of, this paper.  These are:

\begin{enumerate}

\item All calculations in this paper, as well as those in previous analyses of the $B-L$ MSSM such as \cite{Dumitru:2018nct}, are carried out using the {\it one-loop 
corrected} $\beta$ and $\gamma$ renormalization group functions associated with the dimensionless and dimensionful parameters of the theory. However, we systematically {\it ignore all higher-loop corrections to the RGEs as well as any finite one-loop and higher-loop corrections} to the effective Lagrangian. For the purposes of this paper this is sufficient, since our goal is to present the allowed RPV decay channels of the Bino neutralino LSPs in the $B-L$ MSSM theory and to give their {\it leading order} decay rates, branching ratios and the relationship of these to the neutrino mass hierarchy. 
However, the calculations presented here could be expanded to higher precision--that is, to finite one-loop and higher-loop RG/finite corrections--using computational formalisms such as in ISAJET \cite{Paige:2003mg}, FlexibleSUSY \cite{Athron:2014yba},
NMSPEC  \cite{Ellwanger:2006rn}, SUSPECT \cite{Djouadi:2002ze}, SARAH \cite{Staub:2008uz}, SPHENO \cite{Porod:2003um}, SUSEFLAV \cite{Chowdhury:2011zr} and the latest version of SOFTSUSY \cite {Allanach:2016rxd}. This would put the $B-L$ MSSM computations on the same footing as the the more commonly studied MSSM. We will carry out these higher-loop RG and finite corrections to the $B-L$ MSSM in future publications.

\item In this paper, as well as our previous papers \cite{Ovrut:2014rba, Ovrut:2015uea,Dumitru:2018jyb,Dumitru:2018nct}, the initial soft supersymmetry breaking parameters are selected statistically using a {\it ``log-uniform''} distribution over a mass range compatible with LHC energies. As discussed Section 2 of this paper, this is the {\it standard} distribution used in analyzing  such initial conditions. We are aware that one could choose other statistical distributions for the initial parameters--such as a uniform distribution. However,
for the reasons discussed in detail in \cite{Dumitru:2018nct}, the log-uniform distribution is sufficient for the purposes of this paper. Furthermore, a complete description of initial distributions over the soft SUSY breaking parameters would require an analysis of the explicit mechanism for spontaneous supersymmetry breaking--which is beyond the scope of the present paper.
 We refer the reader to \cite{Dumitru:2018nct} for details.

\item Finally, there is a long literature discussing RPV decays within a vast variety of contexts. Reference \cite{Barbier:2004ez} reviews the theoretical aspects of RPV violation with both bilinear and trilinear RPV couplings added in the superpotential. Relevant to the content of our present paper, this review discussed both explicit and, more briefly, spontaneous RPV due to both left- and right-chiral sneutrinos developing VEVs. More recently, the subject was reviewed in 2015 \cite{Mohapatra:2015fua}. This discussed explicit RPV in the MSSM but, in particular, focused on spontaneous breaking of $R$-parity in theories where the standard model symmetry is extended by a gauged $U(1)_{B-L}$.  More recently, there was a comprehensive paper \cite{Dercks:2017lfq} 
investigating the phenomenology of the MSSM extended by a single trilinear RPV coupling
at the unification scale. It goes on to discuss the RPV decay of some of the LSPs; specifically the Bino neutralino and the stau sparticle, within the context of the RPV-CMSSM. The mechanism of generating Majorana neutrino masses through RPV bilinear terms is treated in \cite {Hirsch:2008ur, Kayla2013, Mitsou:2015eka, Mitsou:2015kpa}. This set of papers also studies the decay modes of some LSPs, with emphasis on the decay modes of the lightest neutralino. There are papers such as \cite{Bomark:2014rra, Csaki:2015uza, Dercks:2017lfq}, which study the RPV decay signatures of chargino, stop, gluinos and charged and neutral Higgsinos, using parameter scans in agreement with the existent experimental bounds. However, they work in different, more general theoretical contexts than our own.

The RPV decays of the Bino neutralino LSPs presented in this paper share many of the concepts and techniques contained in these papers, such as RG evolution, the associated LSP calculations and their RPV decays, relationship to neutrino masses and so on. However, the purpose of our present paper is to discuss the RPV decays of Bino neutralino LSPs precisely within the context of the $B-L$ MSSM; a minimal and specific extension of the MSSM with spontaneously broken $R$-parity. Furthermore, the initial conditions of this theory are chosen so as to be {\it completely consistent with all phenomenological requirements}, a property not shared by much of the previous literature. Our analysis is performed so as to predict RPV LSP decays amenable to observation at the LHC and arising from a minimal, realistic, $N=1$ supersymmetric theory. The calculation of the leading order RPV decays of the Bino  neutralino LSPs in this specific context have not previously appeared in the literature.

\end{enumerate}

\section{Physically Acceptable Vacua}

The soft SUSY breaking Lagrangian has over 100 independent parameters, including gaugino mass terms, sfermions mass terms and Higgs couplings. Imposing physical symmetry constraints, however, the number of independent parameters is reduced to only 24. We refer the reader to \cite{Ovrut:2015uea,Dumitru:2018jyb} for details. However, there is no current experimental data to constrain the scale of the remaining mass terms in the SUSY-breaking Lagrangian, nor their relative sizes.

We choose to statistically scatter the absolute value of all dimensionful soft supersymmetry breaking parameters in the mass interval
\begin{equation}
\big[~\frac{M}{f},Mf~\big] \quad {\rm where}~~~M=1.5~{\rm TeV}~, ~f=6.7 \ .
\label{burt1}
\end{equation}
This guarantees that the absolute value of all mass parameters in the theory lie approximately in the range 
\begin{equation}
\big[200~{\rm GeV},10~{\rm TeV}\big] \ .
\label{burt2}
\end{equation}
As mentioned above, the values of $M$ and $f$ were chosen to maximize the number of points that are of phenomenological interest --- that is, compatible with current LHC bounds while also being potentially amenable to observation at the LHC. Varying $M$ and $f$ change the overall scale of SUSY breaking and the range of values that the soft SUSY breaking parameters can take. However, analysis shows that they do not significantly impact the values of the RPV branching ratios and decay length of the Bino LSP.
The soft supersymmetry breaking parameters are statistically scattered in the range \eqref{burt1} with a log-uniform distribution. This is the {\it standard choice} of prior distribution. For examples and discussion see \cite{Dumitru:2018nct, Athron:2017fxj,Fichet:2012sn,Fundira:2017vip,Bomark:2014rra}. 
Finally, symmetry considerations require the mass parameters in the soft SUSY breaking Lagrangian to be real, while allowing them to take both positive and negative values. Therefore, random ``+'' and ``-'' signs are assigned to each of these parameters.

We statistically generate $10^8$ sets of initial points, each set containing 24 soft SUSY breaking parameters. The absolute value of each soft mass is confined to be in the interval \eqref{burt2} presented above. Using RGE calculations discussed in detail in \cite{Ovrut:2015uea,Dumitru:2018jyb}, we run the mass parameters and physical couplings of the theory down from the scale $M_{I}$--where the $B-L$ MSSM emerges after the spontaneous breaking of an $SO(10)$ GUT theory--through the soft SUSY breaking scale to lower energies. However, not all sets of random throws are physically viable at low energy. We need to ensure that $B-L$ and electroweak symmetries are broken at the right energy scales and produce massive bosons in agreement with the current phenomenological constraints. Presently, the lower bound on the $Z_R$ boson, produced after $B-L$ symmetry breaking, is~\cite{Aaboud:2017buh}
\begin{equation}
M_{Z_R}\geq 4.1 \> \text{TeV}.
\end{equation}
Electroweak (EW) symmetry must be spontaneously broken so that the $Z^{0}$ and $W^{\pm}$ masses have their measured values of \cite{PDG}
\begin{equation}
\quad M_{{Z^{0}}}= 91.1876 \pm 0.0021~{\rm GeV} , \quad M_{W^{\pm}}= 80.379 \pm0.012~{\rm GeV} \ .  \label{eq:40}
\end{equation}
Furthermore, the predicted supersymmetric particles masses must satisfy their current measured lower bounds, given in Table \ref{tab:lower_bounds}.
\begin{table}[t]

\begin{center}

\begin{tabular}{ |c|c| }

\hline

SUSY Particle & Lower Bound \\

\hline

Left-handed sneutrinos & 45.6 GeV\\

Charginos, sleptons& 100 GeV \\

Squarks, except stop or bottom LSP& 1000 GeV \\

Stop LSP (admixture)& 550 GeV \\

Stop LSP (right-handed)& 400 GeV\\

Sbottom LSP& 500 GeV\\

Gluino& 1300 GeV\\

\hline

\end{tabular}

\end{center}
\caption{Current lower bounds on the SUSY particle masses.}
\label{tab:lower_bounds}

\end{table}
Finally, the Higgs mass must be within the $3\sigma$ allowed range from ATLAS combined run 1 and run 2 results \cite{Aaboud:2018wps}. This is found to be
\begin{equation}
M_{h^{0}}=124.97 \pm 0.72~ {\rm GeV} \ .
\label{eq:41}
\end{equation}

\begin{figure}[t]

\centering

\begin{subfigure}[t]{0.7\textwidth}

\includegraphics[width=1.\textwidth]{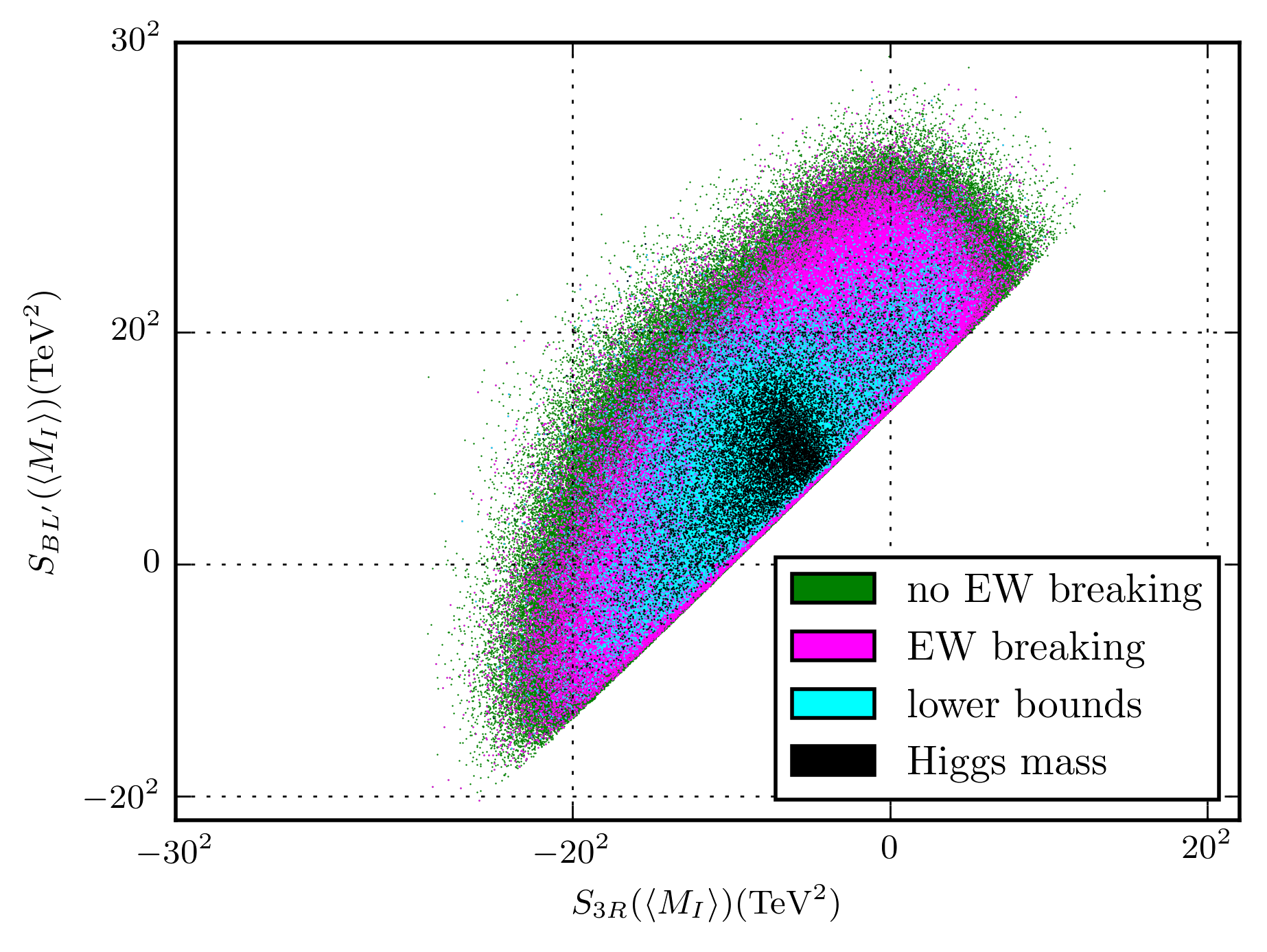}

\end{subfigure}

\caption{Plot of the 100 million initial data points for the RG analysis evaluated at $M_{I}$ . The 4,351,809 green points lead to appropriate breaking of the $B-L$ symmetry. Of these, the 3,142,657 purple points also break the EW symmetry with the correct vector boson masses. The cyan points correspond to 342,236 initial points that, in addition to appropriate $B-L$ and EW breaking, also satisfy all lower bounds on the sparticle masses. Finally, as a subset of these 342,236 initial points, there are 67,576 valid black points which lead to the experimentally measured value of the Higgs boson mass.}\label{fig:ScatterPlot}
\label{fig:eye}
\end{figure}

\begin{figure}[t]
   \centering

   \begin{subfigure}[b]{0.8\textwidth}
\includegraphics[width=1.\textwidth]{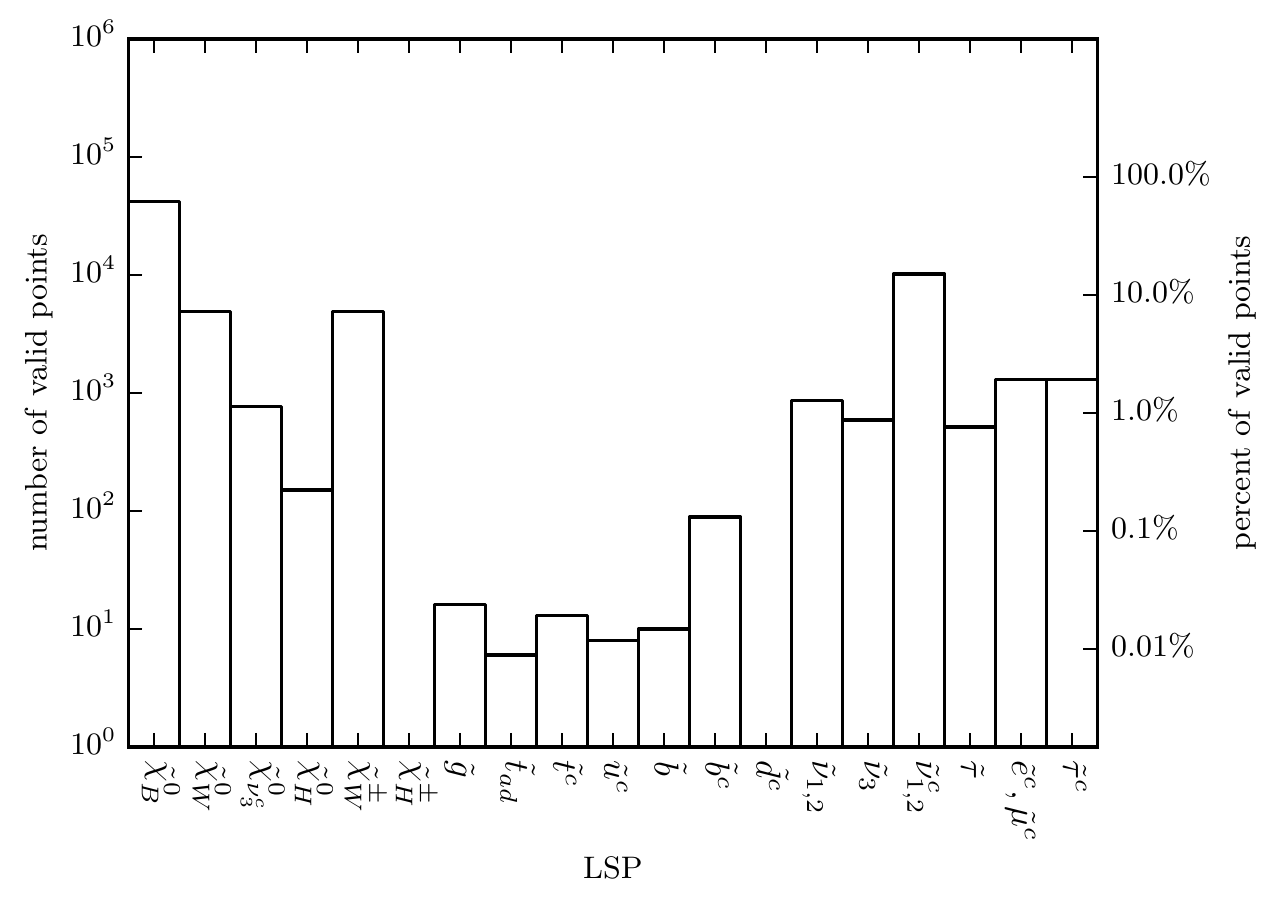}
\end{subfigure}
\caption{A histogram of the LSPs associated with a random scan of 100 million initial data points, showing the percentage of valid black points with a given LSP. Sparticles which did not appear as LSPs are omitted. The y-axis has a log scale. The notation and discussion of the sparticle symbols on the x-axis were presented in \cite{Dumitru:2018jyb}.}
\label{fig:LSP_Hist}
\end{figure}

In the end, out of the 100 million initial sets of parameters, only 65,576 satisfy all the physical requirements above. These sets will be refereed to as ``viable points'' or ``black points'' for the remainder of this paper.  The process of checking the physical  constraints is most clearly exemplified in the 2D scatter plot in Figure \ref{fig:eye}. The number of physically  acceptable points decreases as we enforce, step by step, all of the phenomenological conditions discussed above.
\noindent Each of the physical 65,576 black points found in our simulation represents a set of initial conditions with a distinct low energy sparticle spectrum. The species and mass of the LSP differs from point to point. We show a statistical distribution of the LSPs in the histogram in Figure \ref{fig:LSP_Hist}. It is immediately clear that the most likely LSP candidate is the Bino neutralino $\tilde \chi_B^0$. We find that 42,039 black points out of the 67,576 physically viable sets have a Bino neutralino LSP. Other favorable candidates are the Wino neutralino  $\tilde \chi^0_W$ and the Wino chargino $\tilde \chi^\pm_W$, associated with 4,869 and 4,858 black point respectively, and the right handed sneutrinos $\tilde \nu^c_{1,2}$.
The Wino neutralino  $\tilde \chi^0_W$ and Wino chargino $\tilde \chi^\pm_W$ RPV decays were studied in detail in \cite{Dumitru:2018nct}.

\section{The Bino Neutralino}

\indent In the {\it absence of the RPV violating terms} proportional to $\epsilon_{i}$ and $v_{L_{i}}$,  the neutral Higgsinos and gauginos of the theory mix with
the third generation right handed neutrino. In the gauge eigenstate basis $\psi^0=\left( \tilde{W}_R, \tilde{W}_0, \tilde{H}_d^0,
\tilde{H}_u^0, \tilde{B}^\prime,{\nu}^c_3 \right)$, 
\begin{equation}
\mathcal{L}\supset-\frac{1}{2}\left(\psi^0\right)^T{M}_{\tilde{{ \chi}}^0}\psi^0+c.c
\end{equation}
where 
\begin{equation}
M_{{\tilde \chi}^0}=
\left(
\begin{matrix}
M_R&0&-\frac{1}{2}g_Rv_d&\frac{1}{2}g_Rv_u&0&-\frac{1}{2}g_Rv_R\\
0&M_2&\frac{1}{2}g_2 v_d&-\frac{1}{2}g_2 v_u&0&0\\
-\frac{1}{2}g_Rv_d&\frac{1}{2}g_2 v_d&0&-\mu&0&0\\
\frac{1}{2}g_Rv_u&-\frac{1}{2}g_2 v_u&-\mu&0&0&0\\
0&0&0&0&M_{BL}&\frac{1}{{2}}g_{BL}v_R\\
-\frac{1}{{2}}g_Rv_R&0&0&0&\frac{1}{{2}}g_{BL}v_R&0\\
\end{matrix}
\right) \ .
\label{eq:neutralinoMassMatrixWithoutEpsilon}
\end{equation}
In the neutralino mass mixing matrix shown in \eqref{eq:neutralinoMassMatrixWithoutEpsilon}, $M_2$, $M_R$ and $M_{BL}$ are the gaugino mass terms 
introduced in the soft SUSY breaking Lagrangian. They correspond to the symmetry groups $SU(2)_{L}$, $U(1)_{3R}$ and $U(1)_{B-L}$ respectively, The associated gauge couplings are  $g_2$, $g_R$ and $g_{B-L}$. In our simulation, we sample the absolute values of the gaugino masses between $\big[200~{\rm GeV},10~ {\rm TeV}\big]$ , as discussed in Section 1, and further allow them to have either positive or negative signs, which are determined statistically.
The $\mu$ parameter is the Higgsino mass term. Its value is chosen so as to produce the correct $Z^0$ and $W^{\pm}$ boson masses, a process called the ``little hierarchy problem''  \cite{Ovrut:2015uea}. The dimensionful parameters $v_u$ and $v_d$ are the Higgs up and Higgs down VEVs that break electroweak symmetry, while $v_R$ is the third generation sneutrino VEV, which breaks $B-L$ symmetry at a much higher scale.

The $B-L$ MSSM does not explicitly contain a Bino, associated with the hypercharge group $U(1)_Y$.  Instead, it contains a Blino $\tilde B^'$ and a Rino $W_R$, the gauginos associated with the symmetry groups $U(1)_{B-L}$ and $U(1)_{3R}$, respectively. Nevertheless, the theory does effectively contain a Bino. This is most easily seen using the following approximation. Let us consider the limit $M_{W^\pm}^2,\> M_{Z^0}^2\ll M_{R}^{2}, \> M_{2}^{2},\> M_{BL}^{2},\mu^{2}$ --- that is, when the 
EW scale is much lower than the 
soft SUSY breaking scale so that the Higgs VEV's are negligible. Note that $\mu^{2}$ appears in these inequalities since, as discussed in \cite{Ovrut:2015uea}, it must be on the order of the soft SUSY breaking Higgs parameters $m_{H_{u}}^{2},m_{H_{d}}^{2}$ to solve the ``little hierarchy problem''. In this limit, the mass matrix in eq. \eqref{eq:neutralinoMassMatrixWithoutEpsilon} becomes
\begin{equation}
M_{{\tilde \chi}^0}=
\left(
\begin{matrix}
M_R&0&0&0&0&-\frac{1}{ 2}g_Rv_R\\
0&M_2&0&0&0&0\\
0&0&0&-\mu&0&0\\
0&0&-\mu&0&0&0\\
0&0&0&0&M_{BL}&\frac{1}{{2}}g_{BL}v_R\\
-\frac{1}{{2}}g_Rv_R&0&0&0&\frac{1}{{2}}g_{BL}v_R&0\\
\end{matrix}
\right)
\label{wow1}
\end{equation}
The first, fifth, and sixth columns, corresponding to the Blino, the Rino and the third generation right-handed neutrino, are now decoupled from the other three states and mix only with each other. In the reduced basis $\left({\nu}_3^c, \tilde W_R, \tilde B^\prime \right)$, the off-diagonal mass matrix is
\begin{equation}
\left(
\begin{matrix}
0&-\cos \theta_R M_{Z_R}&\sin \theta_R M_{Z_R}\\
-\cos \theta_R M_{Z_R}&M_R&0\\
\sin \theta_R M_{Z_R}&0&M_{BL}\\
\end{matrix}
\right)
\end{equation}
with
\begin{equation}
M_{Z_{R}}=\frac{1}{2}\sqrt{g_{R}^{2}+g_{BL}^{2}}~ v_{R}~~,~~\cos \theta_R = \frac{g_R}{\sqrt{g_R^2+g_{BL}^2}} \ .
\end{equation}
%
Note that the experimental lower bound on $M_{Z_R}$ is much higher than the typical physical gaugino mass lower bounds. This mass hierarchy is also motivated theoretically because RG running makes the gauginos masses lighter than $M_{Z_R}$; that is, $M^2_{R},M^2_{BL} \ll M^2_{Z_R}$. See  \cite{Ovrut:2015uea} for details. Taking this limit, the mass eigenstates and eigenvalues can be found as an expansion in the gaugino masses. To {\it zeroth order}, the mass eigenstates are
\begin{equation}
{\tilde B}=\tilde W_R \sin \theta_R+\tilde B^\prime \cos \theta_R \ ,
\label{train1}
\end{equation}
\begin{equation}
{\nu}_{3a}^c=\frac{1}{\sqrt 2}({\nu^c}_3-\tilde W_R \cos \theta_R+\tilde B^\prime 
\sin \theta_R) \ ,
\end{equation}
\begin{equation}
{\nu}_{3b}^c=\frac{1}{\sqrt{2}}({\nu^c}_3+\tilde W_R \cos \theta_R-
\tilde B^\prime \sin \theta_R) \ .
\end{equation}
Note that, to leading order, \eqref{train1} defines the ``Bino'' in terms of $W_R $ and $\tilde B^\prime$.
The associated mass eigenvalues, calculated to leading order, are given by
\begin{equation}\label{m1eqn}
M_1=\sin^2 \theta_R M_R+\cos^2 \theta_R M_{BL},
\end{equation}
for the Bino, and
\begin{equation}
 m_{{\nu^c}_{3a}}=M_{Z_R},
\quad m_{{\nu^c}_{3b}}=M_{Z_R} \ .
\end{equation}
for two species of massive right handed neutrinos.

\begin{figure}[t]
   \centering

   \begin{subfigure}[c]{0.49\textwidth}
\includegraphics[width=1.0\textwidth]{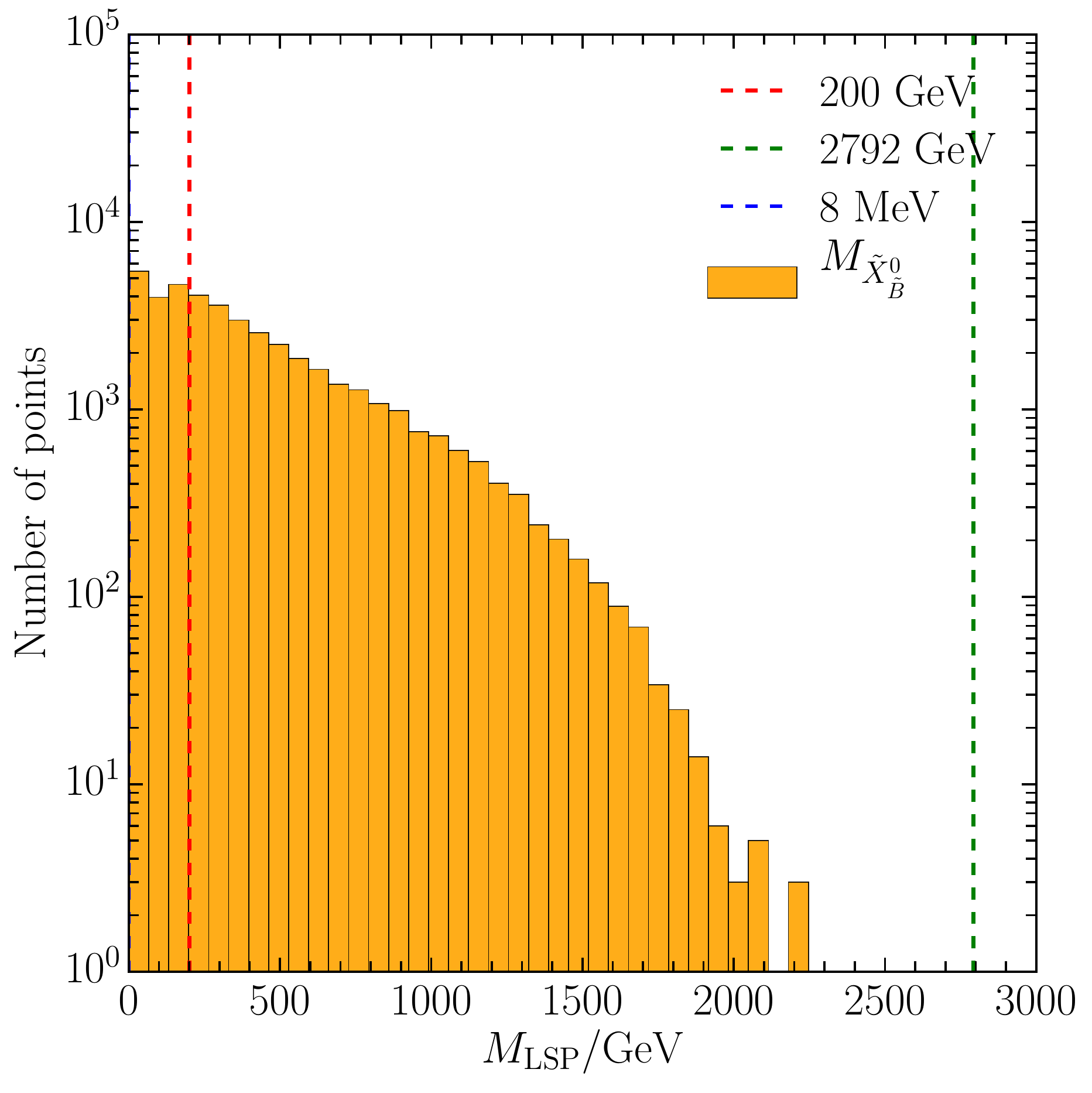}
\caption{}
\label{fig:mass_hist2}
\end{subfigure}
   \begin{subfigure}[c]{0.49\textwidth}
\includegraphics[width=1.0\textwidth]{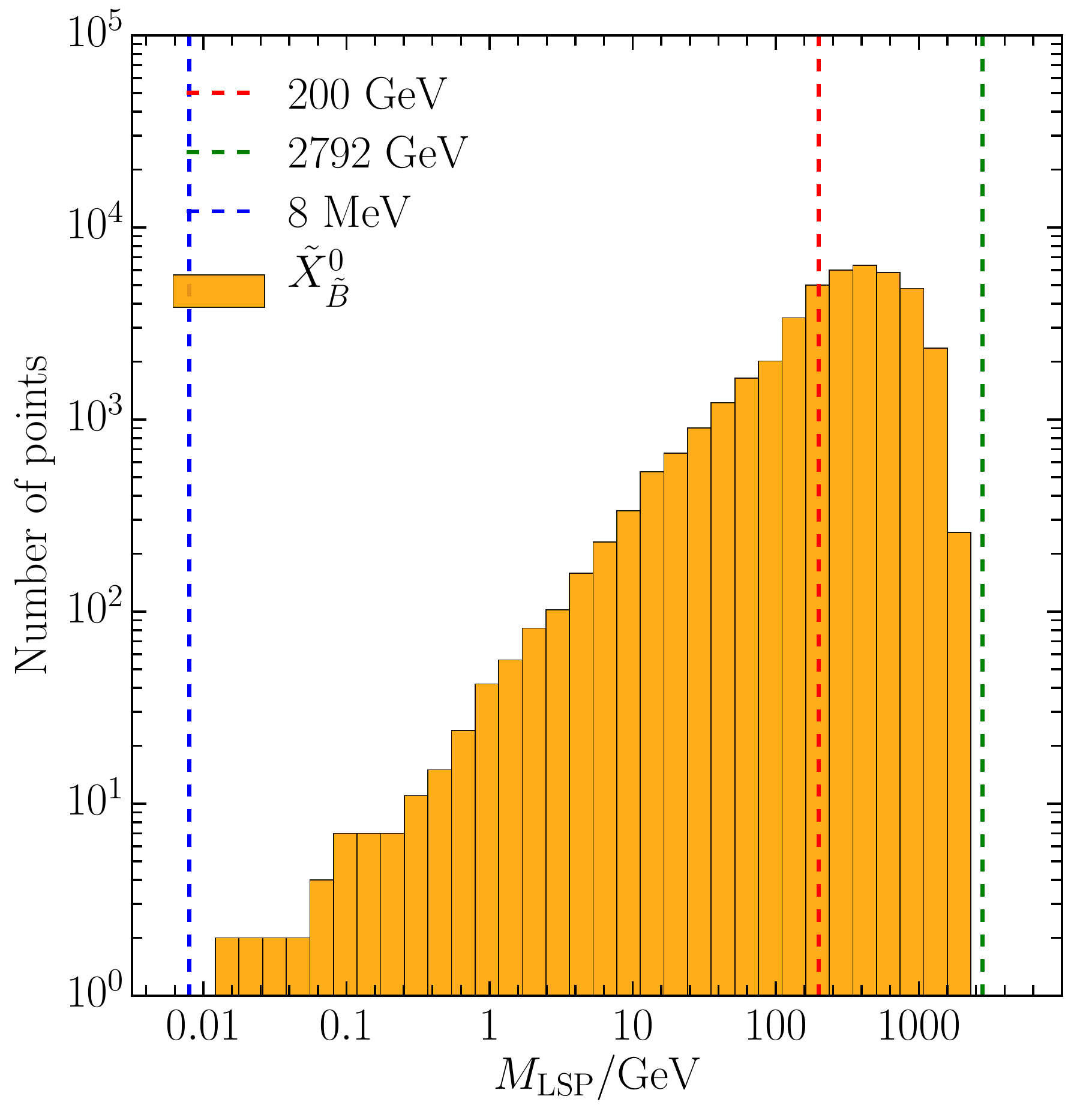}
\caption{}
\label{fig:mass_hist1}
\end{subfigure}
\caption{ The distribution of the Bino neutralino LSP masses for the 42,039 valid black points, shown with linear (a) and logarithmic (b) mass scales. The masses range from 8 MeV to 2792 GeV. Each of the boundary masses occurs only once out of the 42,039 valid points and, hence, they cannot be seen in the histogram.}
\end{figure}

Having defined the Bino, ${\tilde B}$, as well as the ${\nu}_{3a}^c$ , ${\nu}_{3b}^c$ states using the above approximations, we now return to the full gauge basis 
$\psi^0=\left( \tilde{W}_R, \tilde{W}_0, \tilde{H}_d^0,\tilde{H}_u^0, \tilde{B}^\prime,{\nu}^c_3 \right)$ and numerically diagonalize the complete mass matrix $M_{{\tilde \chi}^0}$ given in \eqref{eq:neutralinoMassMatrixWithoutEpsilon}-- without any approximations.
The mass eigenstates are related to the gauge states by the unitary matrix $N$ where ${\tilde{\chi }}^{0}=N \psi^{0}$. $N$ is chosen so that

\begin{equation}\label{eq:neutralino_states}
N^*M_{\tilde{{ \chi}}^0}N^{\dag}= M_{\tilde{{ \chi}}^0}^{D} =
\left(
\begin{matrix}
M_{{\tilde \chi}^0_1}&0&0&0&0&0\\
0&M_{{\tilde \chi}^0_2}&0&0&0&0\\
0&0&M_{{\tilde \chi}^0_3}&0&0&0\\
0&0&0&M_{{\tilde \chi}^0_4}&0&0\\
0&0&0&0&M_{{\tilde \chi}^0_5}&0\\
0&0&0&0&0&M_{{\tilde \chi}^0_6}\\
\end{matrix}
\right) \ ,
\end{equation}
where all eigenvalues are positive.  After diagonalizing the neutralino mass matrix, one obtains six neutralino mass eigenstates, $\tilde \chi_n^0$ with $n=1,2,3,4,5,6$. Unlike for charginos, the label $n$ does not automatically imply any mass ordering; for example, the $\tilde \chi_1^0$ neutralino is not necessarily the lightest. Each of the six neutralinos $\tilde \chi_n^0$ is a superposition of a Rino $\tilde W_R$, a Wino $\tilde W_2$, two neutral Higgsinos $\tilde H_d^0$, $\tilde H_u^0$, a Blino $\tilde B^'$ and a third generation right handed neutrino $\nu_3^c$.  In the theoretical context we work in,  the off-diagonal terms are much smaller than the diagonal ones. This allows one to determine which component dominates in each of the neutralino states $\tilde \chi_n^0$. We find that $\tilde \chi_1^0$ has a dominant Bino $\tilde B$ component, $\tilde \chi_2^0$ has a dominant Wino $\tilde W$ component, $\tilde \chi_{3,4}^0$ have dominant Higgsino $\tilde H_u^0, \> \tilde H_d^0$ components and $\tilde \chi_{5,6}^0$ have a dominant right-handed neutrino $\nu_3^c$ component. Therefore, we use the notation
\begin{equation}
{\tilde \chi}_1^0={\tilde \chi}_B^0,\quad  {\tilde \chi}_2^0={\tilde \chi}_W^0, \quad {\tilde \chi}_3^0={\tilde \chi}_{H_d}^0,
\quad {\tilde \chi}_4^0={\tilde \chi}_{H_u}^0, \quad {\tilde \chi}_5^0={\tilde \chi}_{\nu_{3a}}^0, \quad {\tilde \chi}_6^0={\tilde \chi}_{\nu_{3b}}^0
\end{equation}
to express which component dominates in each neutralino state. Although it is helpful to display the dominant component in each neutralino, in our calculations we use the {\it exact} content of each neutralino state. This is computed numerically, after diagonalizing the neutralino mass mixing matrix. 



Our discussion thus far was carried out in the absence of the RPV couplings, which are central to our theory. These couple the gaugino, Higgsino and the third generation right-handed neutrino states to the three generations of left-handed neutrinos $\nu_1$, $\nu_2$, $\nu_3$. Therefore, with the RPV extension,  the gauge eigenstate basis is enlarged from  $\psi^0=\left( \tilde{W}_R, \tilde{W}_0, \tilde{H}_d^0
\tilde{H}_u^0, \tilde{B}^\prime,{\nu}^c_3 \right)$  to  $\Psi^0=\left( \tilde{W}_R, \tilde{W}_0, \tilde{H}_d^0,
\tilde{H}_u^0, \tilde{B}^\prime,{\nu}^c_3, \nu_1, \nu_2, \nu_3\right)$ and the neutralino mass matrix becomes $9 \times 9$. After diagonalization, the three generations of left-handed neutrinos receive non-zero Majorana masses, a process carefully outlined in \cite{Dumitru:2018jyb,Dumitru:2018nct}. The original six neutralino eigenstates, on the other hand, each receive a negligibly small RPV contribution containing the three left-handed neutrinos. For example, 
the RPV-extended Bino neutralino mass eigenstate $\tilde \chi_B^0$ is now a linear combination of nine gauge eigenstates; three gaugino states, two Higgsino states, a third generation right-handed neutrino and three left-handed neutrinos. The left-handed neutrino contributions to the Bino neutralino eigenstate and mass are negligible, since they have been introduced via small RPV couplings. Therefore, the Bino neutralino continues to generically have a strongly dominant Bino component $\tilde \chi_B^0\simeq {\tilde{B}}$. Expanding, as discussed above, in the limit that $M_{W^\pm}^2,\> M_{Z^0}^2\ll M_{R}^{2}, \> M_{2}^{2},\> M_{BL}^{2},\>\mu^{2}$-- that is, when the 
EW scale is much lower than the 
soft SUSY breaking scale--but now to to {\it first order}, we find that the Bino mass $M_{{\tilde \chi}^0_B}$ is given by
\begin{equation}\label{eq:Bino_mass}
M_{{\tilde \chi}^0_B} \simeq |M_1|-\frac{M_{Z^0}^2\sin^2 \theta_W(M_1+\mu \sin 2\beta)}{\mu^2-M_1^2} \ .
\end{equation}

In the second term in eq. \eqref{eq:Bino_mass}, the mass $\mu$ is of the order of the soft SUSY breaking mass parameters $m_{H_{u}}^{2},m_{H_{d}}^{2}$ to solve the ``little hierarchy problem''. Statistically, this is always much larger than the mass of the $Z^0$ boson in the numerator. 
Therefore, the mass of the Bino neutralino $\chi_B^0$ is approximately equal to $|M_1|$.
As discussed in \cite{Ovrut:2015uea}, the viable black points must satisfy the inequalities $M_{R}^2,M_{BL}^2 \ll M_{Z_R}^2$ to be consistent with low energy data. To lowest order, therefore, expression \eqref{m1eqn} is  a good approximation to the mass $M_{1}$. Generically, therefore, Bino LSP masses are expected to lie in the same the interval as $|M_{R}|$ and $|M_{BL}|$; that is  $\big[200~{\rm GeV}, 10~{\rm TeV}\big]$. That this is {\it generically} the case can be seen in Figure \ref{fig:mass_hist2}. However, there is a {\it very important caveat} to this statement. Note that 
$M_{R}$ and $M_{BL}$ do not enter expression \eqref{m1eqn} for $M_{1}$ as absolute values; that is, the only constraint on these mass terms in \eqref{m1eqn}  is that they be real--however, they can be either positive or negative. This leaves open the possibility, albeit more unlikely, that the terms in eq. \eqref{m1eqn} can almost, or even exactly, cancel. In such cases, one would obtain small Bino mass terms where $|M_1|<M_{W^\pm}$,  and lead to Bino LSP masses smaller than the EW scale. That such cancellations can indeed occur is shown in Figure \ref{fig:mass_hist1}. Note that such light Bino LSPs can only decay via suppressed off-shell processes, leading to interesting experimental signatures at the LHC.
 However, these light Bino LSPs are statistically much less probable. In Figure \ref{fig:mass_hist1} we see that Bino masses much smaller than 200 GeV are less and less likely for the range of scanned parameters. The analysis presented in this paper involved $10^{8}$ initial statistical samples, leading to 42,039 black points with Bino LSPs. For this sample, the smallest and largest Bino masses we find are 8 MeV and 2792 GeV respectively. However, a larger statistical sample can lead to much smaller, and larger, Bino LSP masses.
Note that the existence of very light Bino LSPs is exciting from a cosmological point of view, since it offers a possible dark matter candidate. However, a study of this cosmological scenario is beyond the scope of this paper and will be presented elsewhere.

In contrast, the masses of other neutralino species cannot become arbitrarily small. This is the case of the Wino neutralino and the Wino chargino, for example. Wino neutralinos and Wino charginos have a dominant Wino component. To first order, the masses of these sparticles are equal, given by
\begin{equation}\label{eq:Wino_Neutralino_mass}
M_{{\tilde \chi}^0_W}=M_{{\tilde \chi}^\pm_W} \simeq |M_2|-\frac{M_{W^\pm}^2(M_2+\mu \sin 2\beta)}{\mu^2-M_2^2} \ .
\end{equation}
Including the higher-order terms, the masses split and form almost degenerate pairs. Similarly as in eq. \eqref{eq:Bino_mass}, the second term in eq. \eqref{eq:Wino_Neutralino_mass} is very small compared to the leading term $|M_2|$, because the mass $\mu$ in the denominator must be much larger than the mass $M_{W^\pm}$ in the numerator. Therefore, the masses of the Wino chargino and the Wino neutralino are both approximately equal to $|M_2|$.
The mass of the Wino gaugino $M_2$ is introduced in the soft SUSY-breaking Lagrangian and, hence, we sample its absolute value in the interval $\big[200~{\rm GeV}, 10~{\rm TeV}\big]$. It is, therefore, fixed to be of the order of the SUSY breaking scale. Unlike the Bino gaugino mass term $M_1$,  soft mass parameter $M_2$ cannot get arbitrarily small. Hence, the masses of the Wino neutralinos and Wino charginos cannot get lower than the EW scale.

Finally, we note that even though the RPV left-handed neutrino components have a negligible contribution to the Bino neutralino eigenstate and mass, they remain central to our study of the RPV decays of the Bino neutralino regardless of its mass. From now on, we will use the 4-component spinor notation for the Bino neutralino state, which in terms of the Bino neutralino Weyl spinor, $\tilde \chi_B^0$ is given by
\begin{equation}
\tilde X^0_B=
\left(
\begin{matrix}
\tilde \chi_B^0\\
\tilde \chi_B^{0 \dag}
\end{matrix}
\right).
\end{equation}

\section{Bino Neutralino LSP RPV Decays with On-Shell  $W^\pm, Z^0, h^0$ Bosons }

\begin{figure}[t]
 \begin{minipage}{1.0\textwidth}
     \centering
   \begin{subfigure}[b]{0.31\linewidth}
   \centering
\includegraphics[width=0.8\textwidth]{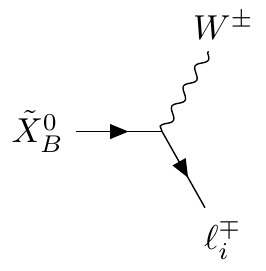}

\caption*{${\tilde X}^0_B\rightarrow W^\pm \ell^\mp_{i}$}
       \label{fig:table2}
   \end{subfigure} 
   \centering
   \begin{subfigure}[b]{0.31\linewidth}
   \centering
\includegraphics[width=0.8\textwidth]{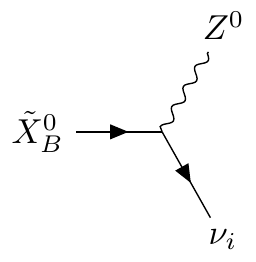}
 
\caption*{ ${\tilde X}^0_B\rightarrow Z^0 \nu_{i}$}
       \label{fig:table2}
\end{subfigure}
   \centering
     \begin{subfigure}[b]{0.31\textwidth}
   \centering
\includegraphics[width=0.8\textwidth]{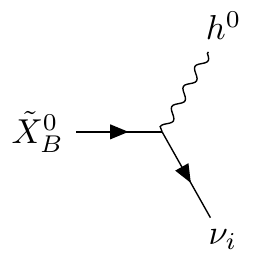}

\caption*{ ${\tilde X}^0_B\rightarrow h^0 \nu_{i}$}
\end{subfigure}
\end{minipage}
\caption{RPV decays of a general massive Bino neutralino $\tilde X_B^0$. There are three possible channels, each with $i=1,2,3$, that allow for Bino neutralino LSP decays. The decay rates into each individual channel were calculated analytically in our previous paper and are reproduced in Appendix \ref{appendix:B}.}\label{fig:NeutralinoDecays}
\end{figure}

We begin by studying the RPV decays of a Bino LSP with {\it mass greater than the electroweak scale} to standard model particles. In the $B-L$ MSSM model, such Bino neutralino LSPs can only have RPV decays into an on-shell massive boson and a lepton. The three possible decay channels, $\tilde X^0_B\rightarrow W^\pm \ell_i^\mp$,  $\tilde X^0_B\rightarrow Z^0 \nu_i$,  $\tilde X^0_B\rightarrow h^0 \nu_i$ for $i=1,2,3$, are shown in Figure \ref{fig:NeutralinoDecays}. Note, however, that all of these decay channels become forbidden at tree level if the mass of the Bino LSP is smaller than the mass of the lightest of the three boson species; that is, the $W^\pm$. This will be the subject of the Section 5. 

\subsection{Branching ratios of the decay channels}

 \begin{figure}[t]
   \centering
   \begin{subfigure}[b]{0.99\textwidth}
\includegraphics[width=1.0\textwidth]{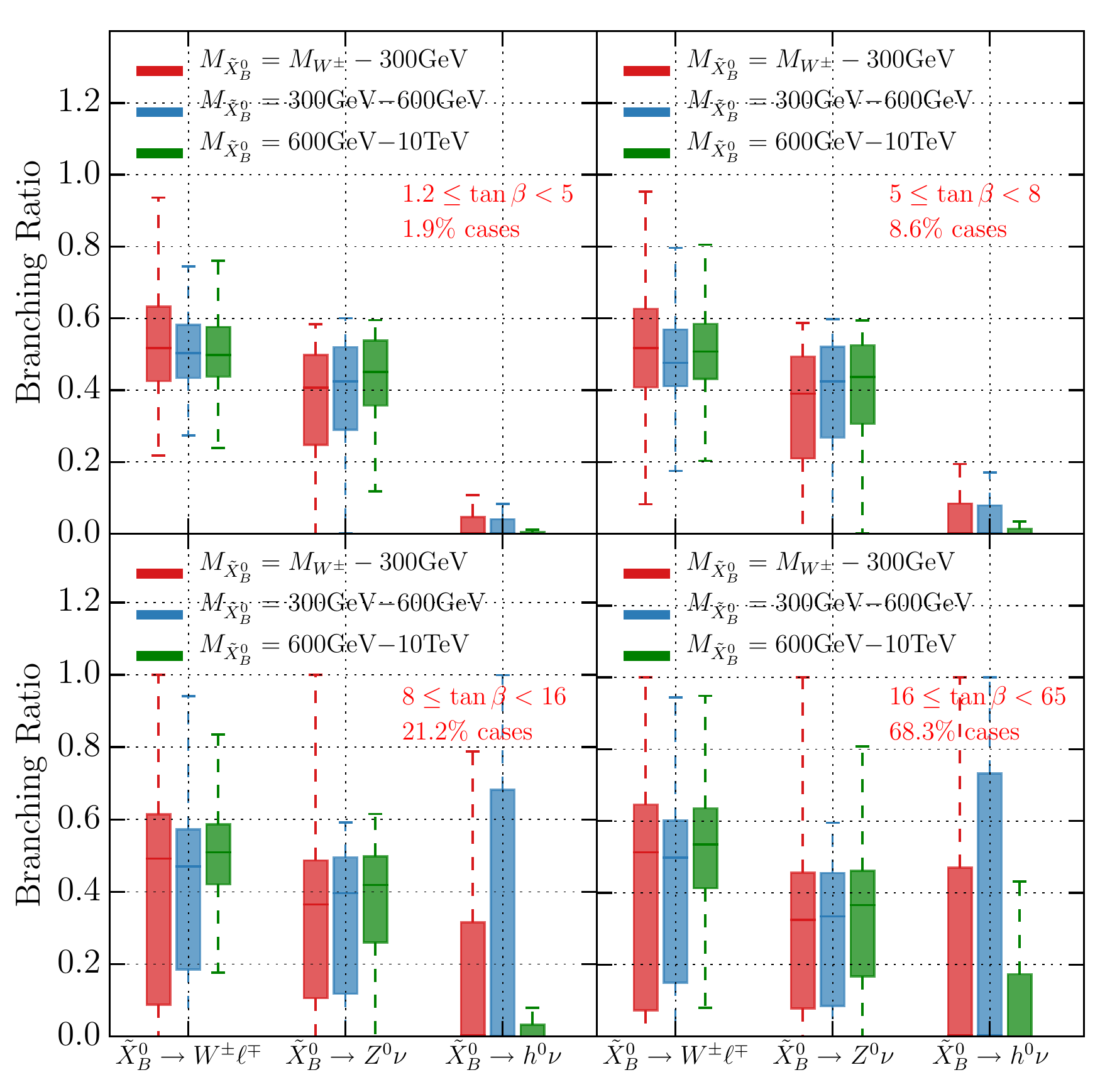}
\end{subfigure}
    \caption{Branching ratios for the three possible decay channels of a Bino neutralino LSP with mass $M_{{\tilde{X}}_{B}^{0}} \geq M_{W^{\pm}}$ divided over three mass bins and four $\tan \beta$ regions. The colored horizontal lines inside the boxes indicate the median values of the branching fraction in each bin, the boxes indicate the interquartile range, while the dashed error bars show the range between the maximum and the minimum values of the branching fractions. The case percentage indicate what percentage of the physical mass spectra have $\tan \beta$ values within the range indicated.  We assumed a normal neutrino hierarchy, with $\theta_{23}=0.597$. Note that the median values of the ${\tilde X}^0_B\rightarrow h^0 \nu$ decay channel approaches zero for all mass ranges and all values of $\tan \beta$.}
\label{fig:bar_plot2}
\end{figure}

In this section, we analyze the RPV decay signatures of Bino neutralino LSPs with masses larger that of the $W^\pm$ boson. We will follow the methods used in our previous  study of Wino chargino and Wino neutralino RPV LSP decays~\cite{Dumitru:2018nct}. Note than in that study, the LSP masses were all found to be at least 200 GeV, so that only decays to on-shell bosons were considered. Of the three Bino decay channels, the  ${\tilde X}^0_B\rightarrow W^\pm \ell^\mp_{i}$
process provides an excellent target for LHC searches, since the final state can be fully reconstructed within the ATLAS detector. In the other two Bino decay processes, the left-handed neutrinos produced via ${\tilde X}^0_B\rightarrow Z^0 \nu_i$ and ${\tilde X}^0_B\rightarrow h^0 \nu_i$ can only be inferred through the presence of missing energy. Hence, the most experimentally clean signature appears to be the Bino neutralino decay into a $W^\pm$ massive boson and a charged lepton.

The relative abundance of each channel is presented in terms of the associated Bino decay branching ratio. For example, for the process ${\tilde X}^0_B\rightarrow W^\pm \ell^\mp$, the branching ratio is defined to be
\begin{equation}
\text{Br}_{{\tilde X}^0_B\rightarrow W^\pm \ell^\mp}=\frac{\sum_{i=1}^{3}  \Gamma_{ {\tilde X}^0_B\rightarrow W^\pm \ell^\mp_{i}}}{\sum_{i=1}^3 \Big( \Gamma_{{\tilde X}^0_B\rightarrow Z^0 \nu_i}+  \Gamma_{{\tilde X}^0_B\rightarrow W^\pm \ell^\mp_i}+\Gamma_{{\tilde X}^0_B\rightarrow h^0 \nu_i}\Big)}  \ ,
\label{fun1}
\end{equation}
where the decay rates, such as  $\Gamma_{ {\tilde X}^0_B\rightarrow W^\pm \ell^\mp_{i}}$, can be constructed from the formulas presented in Appendix B. The expressions for $\text{Br}_{{\tilde X}^0_B\rightarrow Z^0 \nu_i}$ and $\text{Br}_{{\tilde X}^0_B\rightarrow h^0 \nu_i}$ are identical in form to \eqref{fun1}, with the associated decay rates also presented in Appendix B.
In this section, we study the decay patterns and branching ratios for each for the 3 decay channels of the Bino neutralino. As discussed above, there are 42,039 valid black points associated with Bino neutralino LSPs. 
In the present analysis, we retain only the black points with LSPs whose masses are larger than that of the $W^\pm$ bosons.
For each of these, we compute the decay rates via RPV processes, using the expressions \eqref{Neutralino_Decay_Rate1}-\eqref{Neutralino_Decay_Rate3} with $n=1$ given in Appendix \ref{appendix:B}. The branching ratios to each channel take different values for every valid point in our simulation. We compute the median values, interquartile ranges and the minimum and maximum values of the branching fractions 
using the same categories of events as employed in our previous paper for Wino charginos and Wino neutralinos \cite{Dumitru:2018nct}.
Specifically, we sample the average branching fractions in the three bins for the LSP mass $M_{{\tilde X}_B^0} \in [M_{W^\pm}, 300],\>[300,600],\>[600,10^{4}]$~GeV, and in the four intervals for $\tan \beta \in [1.2,5],\> [5,8],\>[8,16], \>[16,65]$. The results are presented in Figure \ref{fig:bar_plot2}. To carry out the explicit calculations, we have chosen a normal neutrino hierarchy with $\theta_{23}=0.597$. We find that assuming an inverted neutrino hierarchy instead produces only minimal changes to these results, while the exact value of $\theta_{23}$ is statistically irrelevant.

It was found--see Figure \ref{fig:bar_plot2}--that the median value of the ${\tilde X}^0_B\rightarrow h^0\nu$ decay channel approaches zero for every mass range and bin for $\tan \beta$. Although the distributions of the branching fractions are fairly broad, we find that they peak very strongly around the median values. It follows that the decay channel is generally  subdominant in all regions of the parameter space. Unlike for the case of Wino charginos and Wino neutralinos, however, we find that $\tan \beta$ has only minimal impact on the experimental predictions.
  While the full expressions for the decay rates are complicated, simplifying assumptions can allow for a better understanding of the relative results.
  One such assumption is that the soft breaking terms have much larger magnitudes than the electroweak scale. This renders the Bino neutralino to be almost purely neutral Bino. Furthermore, the fact that the charged lepton masses are much smaller than the soft breaking parameters further simplifies the equations. Using these approximations in the expressions in Appendix \ref{appendix:B}, one obtains the following simplified formulas for the decay rates. They are given by

\begin{multline}\label{eq:decay_neut1}
\Gamma_{{\tilde X}^0_B\rightarrow Z^0\nu_{i}} \approx
\frac{g_2^2}{16\pi c_W^2}\Big(  
\sin \theta_R 
\Big[ \frac{g_RM_{BL}v_u}{M_1 v_R^2}\epsilon_i +\frac{g_Rg_{BL}^2}{4M_1 \mu}(v_d \epsilon_i+\mu v_{L_i}^*)\Big]\left[ V_{\text{PMNS}}  \right]^\dag_{ij}\\
-\cos \theta_R\Big[\frac{g_R^2g_{BL}}{4M_1 \mu}(v_d \epsilon_i+\mu v_{L_i}^*) - \frac{g_{BL}v_uM_R}{M_1v_R^2}\epsilon_i\Big]\left[ V_{\text{PMNS}}  \right]^\dag_{ij}
\Big)^2
\frac{M_{{\tilde X}_B^0}^3}{M_{Z^0}^2}\left(1-\frac{M_{Z^0}^2}{M_{{\tilde X}_B^0}^2}\right)^2
\left(1+2\frac{M_{Z^0}^2}{M_{{\tilde X}^0_B}^2}\right) \ ,
\end{multline}
\begin{multline}
\Gamma_{{\tilde X}^0_B\rightarrow W^\mp \ell_i^\pm} \approx\frac{g_2^2}{32\pi}\Big(\sin \theta_R 
\Big[ \frac{2g_RM_{BL}v_u}{M_1 v_R^2}\epsilon_i +\frac{g_Rg_{BL}^2}{2M_1 \mu}(v_d \epsilon_i+\mu v_{L_i}^*)\Big]\\
-\cos \theta_R\Big[\frac{g_R^2g_{BL}}{2M_1 \mu}(v_d \epsilon_i+\mu v_{L_i}^*) - \frac{2g_{BL}v_uM_R}{M_1v_R^2}\epsilon_i\Big]
\Big)^2 \times
\frac{M_{{\tilde X}_B^0}^3}{M_{W^\pm}^2}\left(1-\frac{M_{W^\pm}^2}{M_{{\tilde X}_B^0}^2}\right)^2
\left(1+2\frac{M_{W^\pm}^2}{M_{{\tilde X}_B^0}^2}\right) \ ,
\end{multline}
\begin{equation}\label{eq:decay_neut4}
\Gamma_{{\tilde X}^0_B\rightarrow h^0\nu_{i}} \approx\frac{{g_2}^2}{64\pi}\Big( \sin \alpha (\cos^2 \theta_R-\sin^2 \theta_R)\left[ V_{\text{PMNS}}  \right]^\dag_{ij} \frac{\epsilon^*_j }{\mu} \Big)^2
M_{{\tilde X}_B^0}\left(1-\frac{M_{h^0}^2}{M_{{\tilde X}_B^0}^2}\right)^2 \ .
\end{equation}
The notation and derivation of these decay rates is outlined in more detail in \cite{Dumitru:2018jyb}. We also refer the reader to Appendix \ref{appendix:notation} for the definitions of all the parameters in these expressions. We learn that the approximate decay rate for the ${\tilde X}^0_B\rightarrow h^0\nu_{i}$ has an effective coupling proportional to $\frac{{g_2}^2}{64\pi}\Big( \sin \alpha (\cos^2 \theta_R-\sin^2 \theta_R)\Big)^2$. In our theory, $\tan \theta_R=g_{BL}/g_R$ is approximately equal to one. Therefore, $\sin \theta_R \approx \cos \theta_R$, which explains why this channel is subdominant in Bino decays. Furthermore, the expressions for the decay rates of the ${\tilde X}^0_B\rightarrow W^\mp \ell_i^\pm$ and the ${\tilde X}^0_B\rightarrow Z^0\nu_{i}$ channels contain terms that do not depend on $v_d=174~ \text{GeV}/(1+\tan \beta)$. Therefore, we do not observe the suppression of these channels for high $\tan \beta$ values, as is the case for the Wino neutralinos decays presented in \cite{Dumitru:2018nct}.

 \begin{figure}[t]
\begin{subfigure}[t]{0.49\textwidth}
\includegraphics[width=1.\textwidth]{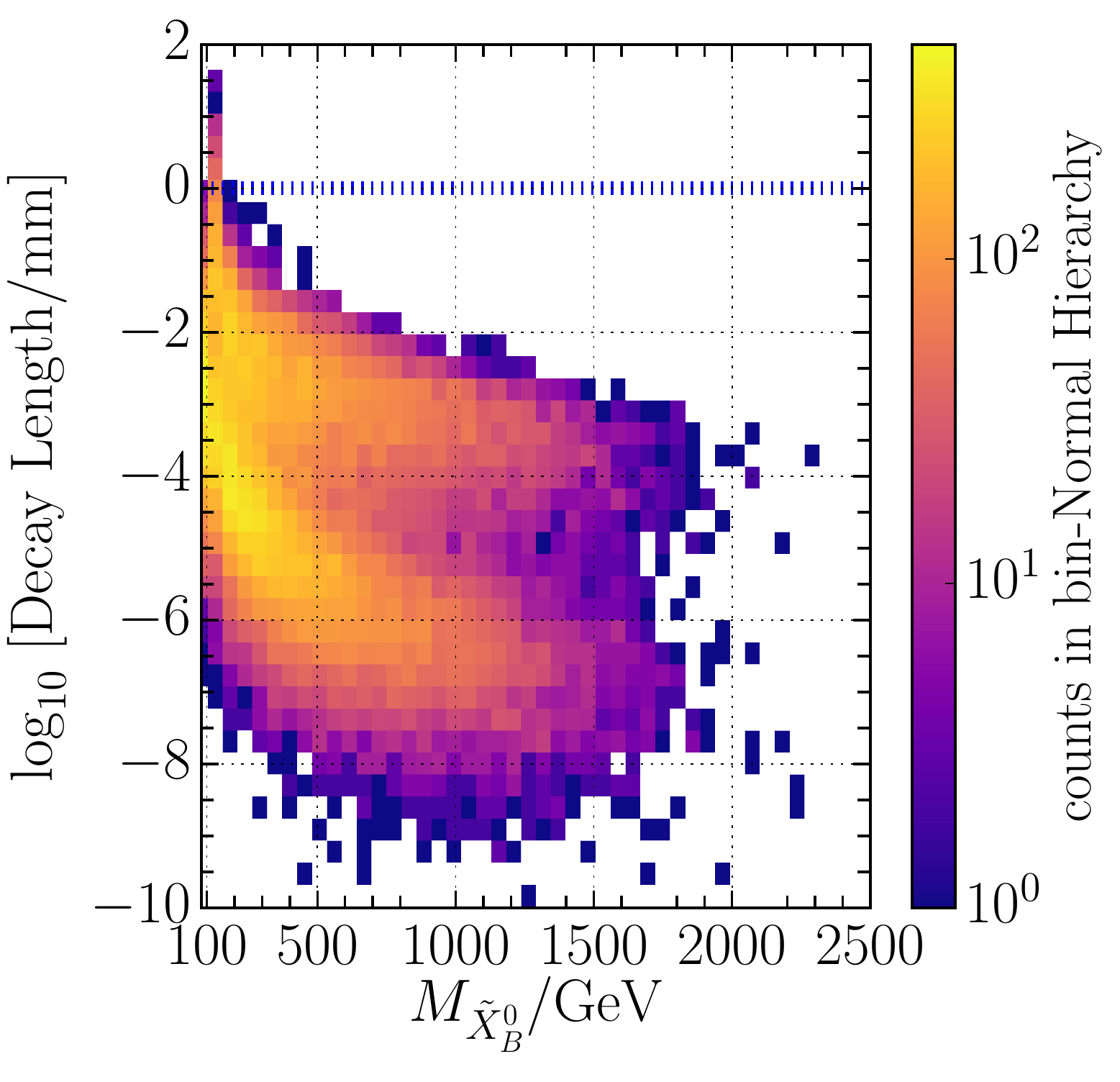}
\end{subfigure}
\begin{subfigure}[b]{0.49\textwidth}
\includegraphics[width=1.\textwidth]{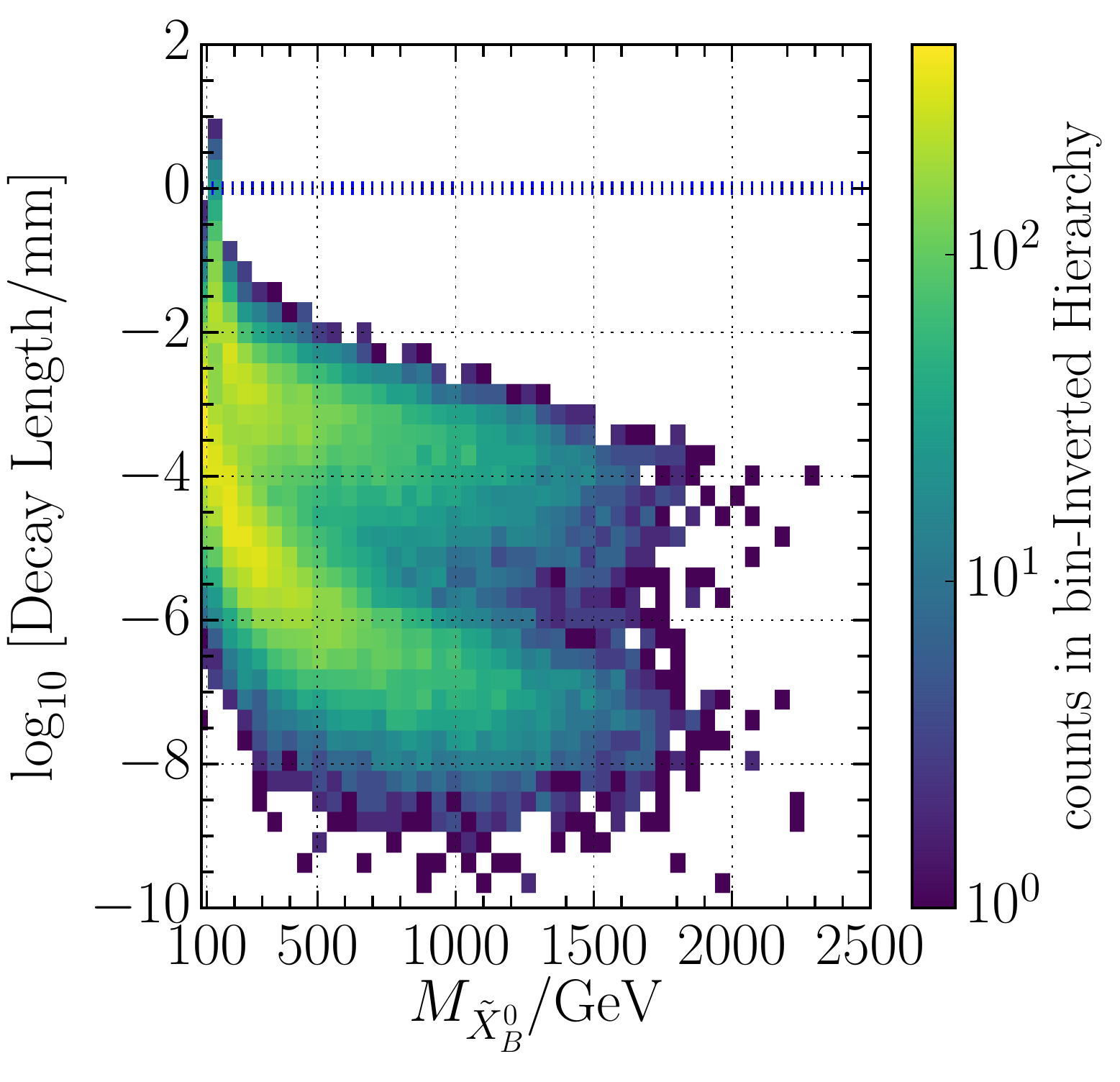}
\end{subfigure}
\caption{Bino neutralino LSP decay length in millimeters, for the normal and inverted hierarchies, summed over the three decay channels. The average decay length $L=c\times \frac{1}{\Gamma}$ decreases for larger values of $M_{\tilde X^0_B}$. We have chosen $\theta_{23}=0.597$ for the normal neutrino hierarchy and $\theta_{23}=0.529$ for the inverted hierarchy. However, the choice of $\theta_{23}$ has no impact on the decay length. The dashed blue line represents the 1~mm decay length, at and below which the decays are ``prompt''. Note that 1) both Figures begin at $M_{W^{\pm}}$ on the left-hand side and 2) there are no points above approximately 2300~GeV. This follows from the fact that, even though the maximum value we obtained for the Bino mass is 2792~GeV, points higher than 2300~Gev are statistically insignificant. See Figure 3. }\label{fig:LSPprompt_neut}
\end{figure}

\subsection{Decay length}

Knowing the branching ratios of the Bino LSP RPV decay channels does not offer a complete picture of the signals that such particle decays can produce in the detector. We further need to analyze the lifetime $\tau$ of these particles by computing of their total decay widths. 
For the purposes of this paper, sparticle decay processes at the LHC are classified into four categories depending on their decay length:

\begin{itemize}
\item{\textit{Prompt decays:} where finite-lifetime effects are experimentally negligible--that is, they do not impact the efficiency of charged lepton reconstruction used by standard analyses. Prompt decays satisfy:~ $c\tau<1$~mm.}
\item{\textit{Displaced vertex} decays: where a \textit{secondary} Bino decay vertex may be identified via charged particle tracking, separate from that of the initial $pp$ interaction. Displaced vertex decays satisfy:~ $1~\text{mm}<c\tau<30$~cm.}

\item{\textit{Decays within the detector:} but outside the tracking apparatus, where measurements made in the muon system may allow observation of the decay. Decays in the detector, but outside the tracking apparatus satisfy:~ $30~\text{cm}<c\tau<10$~m.}

\item{\textit{Detector-stable decays:} where the lifetime is long enough that the only detector signature of the particle is momentum imbalance, that is, ``missing energy''.  Detector-stable decays satisfy:~ $10~m<c\tau$.}

\end{itemize}
\begin{figure}[t]
\centering
\begin{subfigure}[b]{1.\textwidth}
\includegraphics[width=1.0\textwidth]{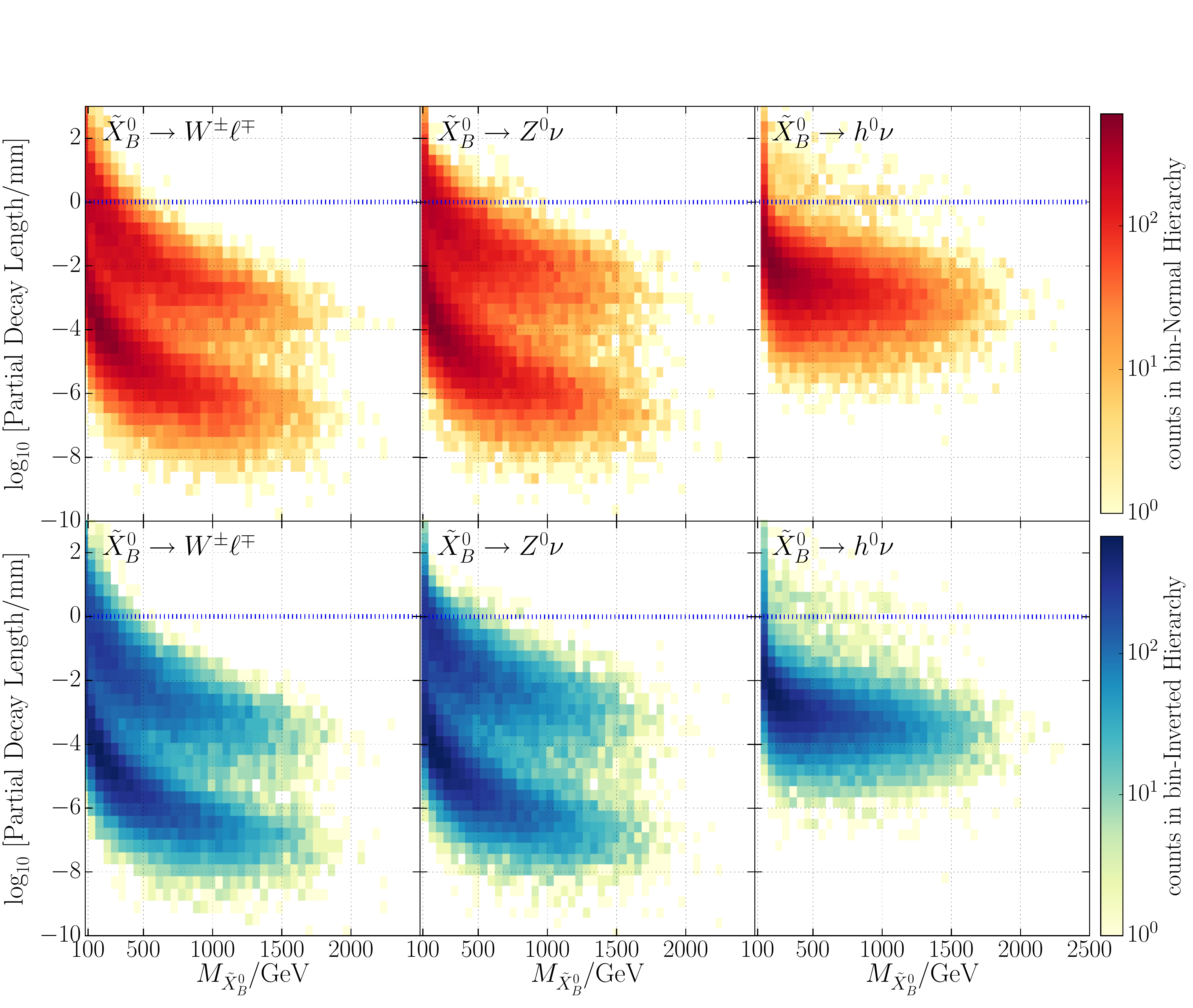}
\end{subfigure}
\caption{Bino neutralino LSP partial decay length	
	in millimeters, shown for the individual decay channels, for both normal and inverted hierarchies. We have chosen $\theta_{23}=0.597$ for the normal neutrino hierarchy and $\theta_{23}=0.529$ for the inverted hierarchy. The choice of $\theta_{23}$ has no impact on the decay length. The blue dashed line denotes a decay length of 1~mm, at and below which decays are ``prompt''. Note that 1) all Figures begin at $M_{W^{\pm}}$ on the left-hand side and 2) there are no points above approximately 2300~GeV. This follows from the fact that, even though the maximum value we obtained for the Bino mass is 2792~GeV, points higher than 2300~GeV are statistically insignificant. See Figure 3.\label{fig:LSPprompt_neut}
}
\label{fig:LSPprompt}
\end{figure}
In reality, search strategies focusing on each of these four cases overlap in sensitivity, partly due to the probabilistic variation in lifetimes of the individual particles produced in $pp$ collisions.
Dedicated searches for long-lived particles have recently been conducted by the ATLAS and CMS Collaborations, searching for displaced charged-particle vertices~\cite{Aaboud:2017iio,Sirunyan:2018vlw}, displaced charged-lepton pairs~\cite{Aad:2015rba,CMS:2014hka}, and displaced jets decaying in the ATLAS muon spectrometer~\cite{Aaboud:2018aqj}.
ATLAS has also studied the complementarity of searches targeting the production of promptly decaying, long-lived, and stable BSM particles to models predicting a wide range of lifetimes~\cite{ATLAS-CONF-2018-003}.
However, the simplified categorization presented above is sufficient for our analysis, indicating the most promising approaches for searching for long-lived Bino LSPs in a given mass range.

%
%

\begin{figure}[t]
   \centering

   \begin{subfigure}[b]{0.44\textwidth}
\includegraphics[width=1.0\textwidth]{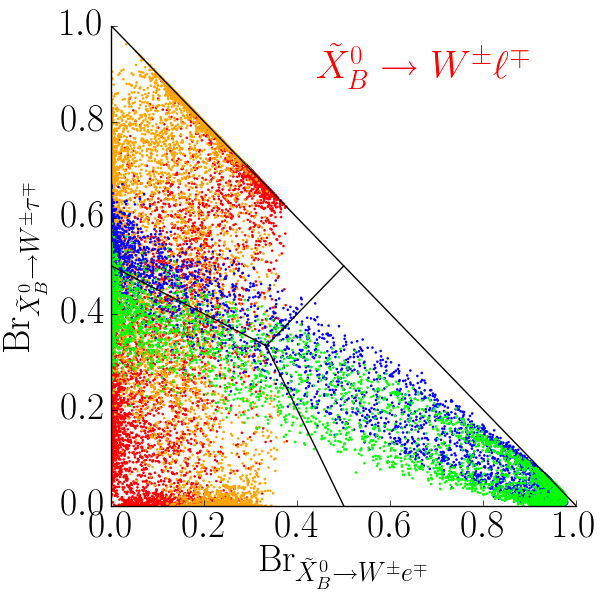}
\end{subfigure}
   \begin{subfigure}[b]{0.44\textwidth}
\includegraphics[width=1.0\textwidth]{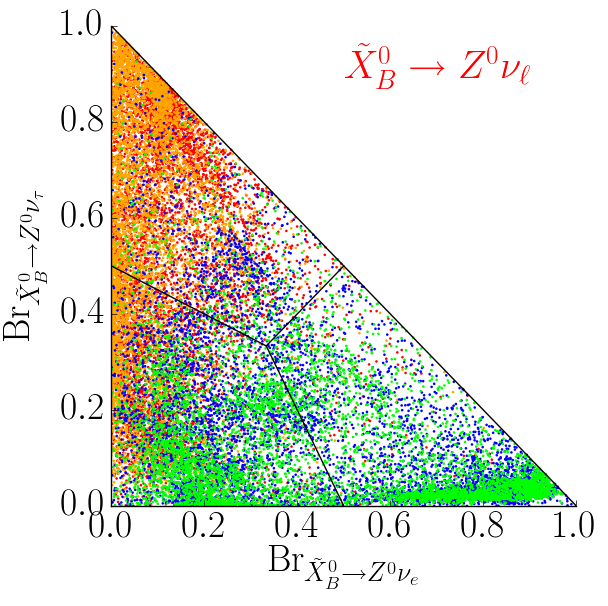}
\end{subfigure}\\
   \begin{subfigure}[b]{0.44\textwidth}
\includegraphics[width=1.0\textwidth]{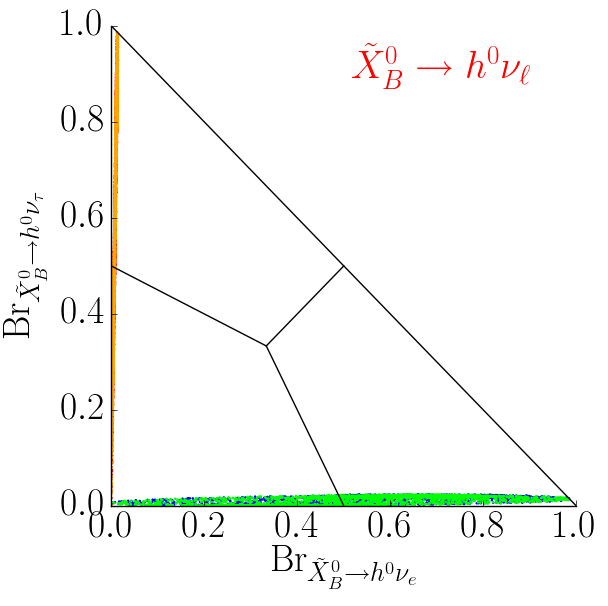} 
\end{subfigure}
   \begin{subfigure}[b]{0.44\textwidth}
\includegraphics[width=1.0\textwidth]{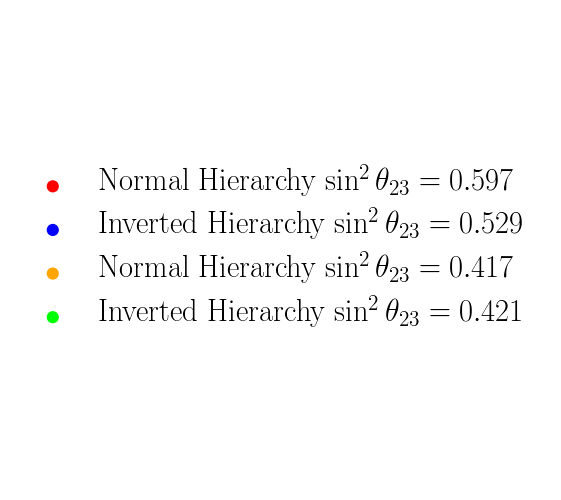}
\end{subfigure}
    \caption{Branching ratios into the three lepton families,
for each of the three main decay channels of a Bino neutralino LSP. The associated neutrino hierarchy and the value of $\theta_{23}$ is specified by the color of the associated data point.}\label{fig:neutralino_lepton_family}
\end{figure}

Figure \ref{fig:LSPprompt_neut} shows that Bino neutralino LSP RPV decays are generally prompt,
with the vast majority of decay lengths found to be less than 1~mm.
Therefore, the Bino decay products may be identified in conventional collider searches without the need for specialized experimental techniques.
We observe that in the case of the inverted hierarchy, the decay lengths are generally smaller, since the values of the RPV couplings are somewhat larger, as explained above.
In Figure \ref{fig:LSPprompt}, the partial decay length
for each of the three decay channels is shown separately.
We find that the total decay width\footnote{The partial decay length is defined as the reciprocal of the partial width for a given decay mode of the Bino.  While results are shown separately for each channel to provide maximal information, the total decay length of the Bino is obtained by combining results across all possible decay modes.} is generally dominated by the  ${\tilde X}^0_B\rightarrow W^\pm \ell^\mp$ and ${\tilde X}^0_B\rightarrow Z^0 \nu$ processes.

\subsection{Lepton family production}

In this subsection, we study the correlation between the electroweak boson and the lepton {\it family} emitted in each of the possible Bino decays.
For example, to quantify the probability to observe an electron $e^\mp$ in the ${\tilde X}^0_B\rightarrow W^\pm \ell^\mp$ process, over a muon $\mu^\mp$ or a tauon  $\tau^\mp$, we compute the relative branching fraction
\begin{equation}
\text{Br}_{{\tilde X}^0_B\rightarrow W^\pm e^\mp}=\frac{\Gamma_{{\tilde X}^0_B\rightarrow W^\pm e^\mp}}{\Gamma_{{\tilde X}^0_B\rightarrow W^\pm e^\mp}+\Gamma_{{\tilde X}^0_B\rightarrow W^\pm \mu^\mp} + \Gamma_{{\tilde X}^0_B\rightarrow W^\pm \tau^\mp} } \ .
\end{equation}

\noindent Using this formalism, we proceed to quantify the branching ratios for each of the three decay processes ${\tilde X}^0_B\rightarrow W^{\pm} \ell^{\mp}$, ${\tilde X}^0_B\rightarrow Z^0 \nu_i$ and ${\tilde X}^0_B\rightarrow h^0 \nu_i$ into their individual lepton families. The results are shown in Figure \ref{fig:neutralino_lepton_family}.
We observe that the ${\tilde X}^0_B\rightarrow W^\pm \ell^\mp$ process has an almost identical statistical distribution for lepton family production as does the Wino chargino decay channel ${\tilde X}^\pm_W\rightarrow Z^0 \ell^\pm$ presented in \cite{Dumitru:2018nct}. Additionally, note that in a Bino neutralino decay via ${\tilde X}^0_B\rightarrow h^0 \nu_i$, the decay rate, given in \eqref{eq:decay_neut4}, has a dominant term proportional to the square of $[V_{\text{PMNS}}^\dag]_{ij}\epsilon_j$. This combination leads to a branching ratio distribution where  no $\nu_\tau$ neutrino is produced in the case of an inverted hierarchy and no $\nu_e$ is produced in the case of a normal hierarchy.

\section{Bino Neutralino LSP RPV Decays with Off-Shell $W^{\pm}, Z^0, h^0$ Bosons }

\begin{figure}[t]
 \begin{minipage}{1.0\textwidth}
     \centering
   \begin{subfigure}[b]{0.49\linewidth}
   \centering
\includegraphics[width=0.8\textwidth]{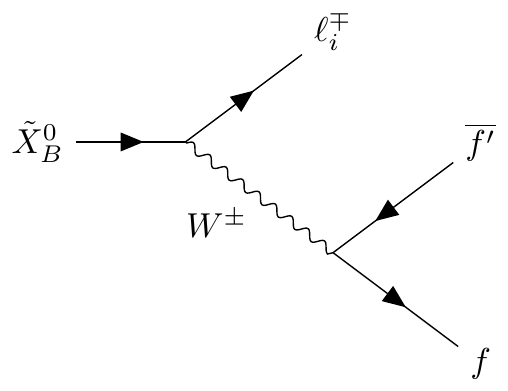}

\caption*{${\tilde X}^0_B\xrightarrow{W^\pm} \ell^\mp_{i}\overline {f^'} f$}
       \label{fig:table2}
   \end{subfigure} \\
   \centering
   \begin{subfigure}[b]{0.48\linewidth}
   \centering
\includegraphics[width=0.8\textwidth]{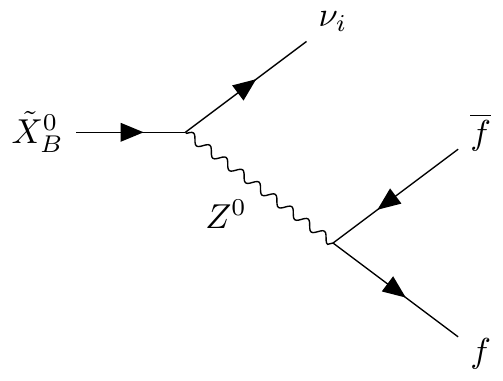}   

\caption*{ ${\tilde X}^0_B\xrightarrow{Z^0} \nu_{i}\overline f f$}
       \label{fig:table2}
\end{subfigure}
   \centering
     \begin{subfigure}[b]{0.48\textwidth}
   \centering
\includegraphics[width=0.8\textwidth]{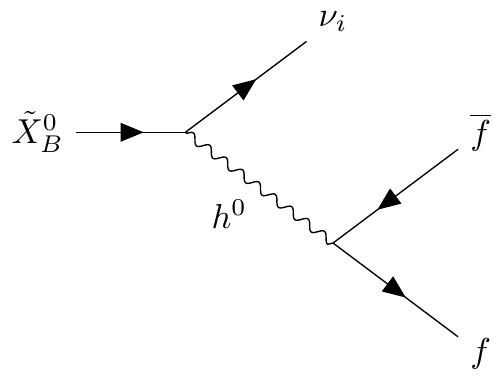}
\caption*{ ${\tilde X}^0_B\xrightarrow{h^0} \nu_{i}\overline f f$}
\end{subfigure}
\end{minipage}
\caption{RPV decays of an $M_{{\tilde{X}}_{B}^{0}} < M_{W^{\pm}}$ Bino neutralino $\tilde X_B^0$ via off-shell $W^\pm$, $Z^0$ and $h^{0}$ bosons, each with three possible channels $i=1,2,3$.
The $W^\pm$ and $Z^0$ bosons may decay to fermion-antifermion pairs, while bosonic decays are also possible in the case of the Higgs. In these Figures, ${\it f^{\prime}}$ represents a generic fermionic state, whereas {\it f} represents a possible fermion or boson decay product. }\label{fig:BinoDecays2}
\end{figure}

In Figures \ref{fig:mass_hist2} and \ref{fig:mass_hist1}, we found that the mass of the Bino neutralino LSP can be as low as a few MeV. For small enough masses, the Bino neutralino LSP can no longer decay via the emission of an on-shell boson as shown in Figure \ref{fig:NeutralinoDecays}. For example,  the process $\tilde X^0_B \rightarrow W^\pm \ell^\mp$ is forbidden if the mass of the Bino neutralino LSP is smaller than the total mass of the $W^\pm$ boson and the accompanying charged lepton. Similarly,  the processes $\tilde X^0_B \rightarrow Z^0 \nu$  and $\tilde X^0_B \rightarrow h^0 \nu$ cannot take place for Bino neutralino LSPs lighter than the $Z^0$ and $h^0$ bosons, respectively. However, in such cases the Bino neutralino LSP will still decay via the RPV processes illustrated in Figure \ref{fig:BinoDecays2}, with intermediate, {\it off-shell} $W^\pm, Z^0$ and $h^0$ bosons.

\subsection{Calculation of off-shell decay widths}

For Binos lighter than $M_{W^{\pm}}$, the processes displayed in Figure \ref{fig:BinoDecays2} are similar to the familiar case of muon decay $\mu\to e \overline{\nu_e} \nu_\mu$.
In muon decay, because the momentum transfer is much smaller than the $W^\pm$ mass, the process may be approximated as an effective 4-point interaction, so that the computation of the decay width becomes straightforward.
In our case, however, the mass of the incoming Bino neutralino LSP is close enough in magnitude to the mass of the off-shell $W^\pm, Z^0$ or $h^0$ bosons that the low-momentum approximation is not generically valid. In general, the decay rate $\Gamma$ is proportional to the coupling strength associated with each vertex, in addition to some dependence on the momentum transfer and the masses of the interacting particles. These contributions can be factorized. To see this, let us consider another sample process; ${\tilde X}^0_B\rightarrow \nu_{i}Z^0$~,~$Z^0 \to\overline f f$. For this process, the decay rate takes the form
\begin{equation}\label{eq:Gamma_form}
\Gamma_{{\tilde X}^0_B  \xrightarrow{Z^0} \nu_{i}\overline f f}=g_{{\tilde X}^0_B\rightarrow Z^0 \nu_{i}}^2g_f^2 F(M_{\tilde X^0_B}, M_{Z^0}, m_f, m_{\nu_i}),
\end{equation}
where $g_{{\tilde X}^0_B\rightarrow Z^0 \nu_{i}}$ is the RPV coupling from the Bino decay vertex,  $g_f$ is the coupling of the $Z^{0}$  boson to $\bar{f}$,$f$ after EW breaking, and $F(M_{\tilde X^0_B}, M_{Z^0}, m_f, m_{\nu_i})$ is a function that only depends on the masses of the particles involved and the width of the intermediate $Z^0$ boson. The expression is obtained after integrating over the momenta of the final particle states. For muon decay, the  mass can be ignored so that $F_{\mu \text{ decay}}$ depends only on the $\mu$ mass and is given analytically exactly by
\begin{equation}
F_{\mu \text{ decay}}=\frac{1}{192\pi}\frac{1}{{M_W^\pm}^4}m_{\mu}^5.
\end{equation}
However, for the Bino decay channels, such as in \eqref{eq:Gamma_form}, analytical calculations of the $F$ functions are non-trivial. Therefore, in this paper, we will compute these functions numerically.

\begin{figure}[t]
 \begin{minipage}{1.0\textwidth}
     \centering
   \begin{subfigure}[b]{0.48\linewidth}
   \centering

\includegraphics[width=0.8\textwidth]{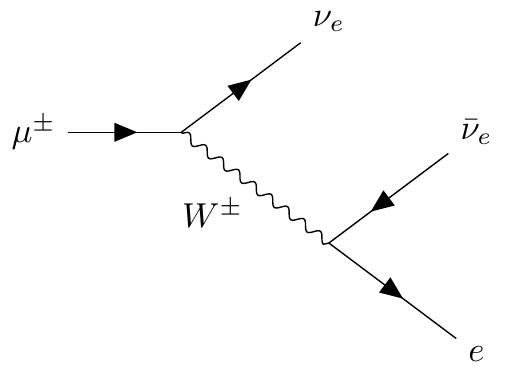}
\caption{ $\mu^\pm \xrightarrow{W^\pm} \nu_{e}\overline{\nu}_e e$}
\label{fig:mudecay}
\end{subfigure}
   \begin{subfigure}[b]{0.48\linewidth}
   \centering
\includegraphics[width=0.8\textwidth]{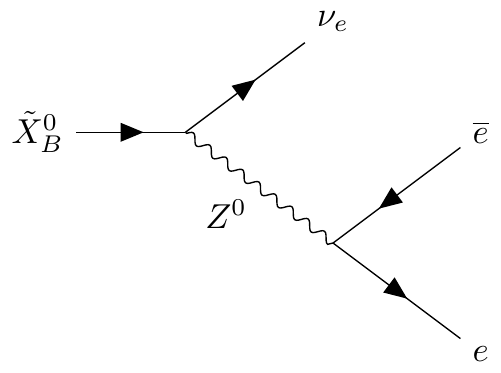}
\caption{${\tilde X}^0_B\xrightarrow{Z^0} \nu_e \overline {e} e$}
       \label{fig:BinoExample}
   \end{subfigure} \\
\end{minipage}
\caption{The Feynman diagram of a $\mu^\pm$ lepton decay (left) has a similar structure to that of a Bino LSP decay (right). We use this similarity to compute the decay rates of a Bino LSP via off-shell bosons, using the Madgraph software.}\label{fig:Madgraph}
\end{figure}

The mass dependence of the Bino decays are calculated using the \textsc{Madgraph5\_aMC@NLO 2.6.4} Monte Carlo event generation program (\madgraph)~\cite{madgraph}, to leading order accuracy in the QCD coupling constant.
Standard Model particle masses and widths are configurable parameters, allowing the kinematic dependence of Bino decay widths to be extracted from a modified calculation of the SM muon decay process $\mu\to e \overline{\nu}_{e} \nu_\mu$ shown in Figure \ref{fig:mudecay}.
The particle test masses in the \madgraph\ calculation (denoted, henceforth, as $\hat m_{\mu}$,$\hat m_{\nu_\mu}$, and so on) can be set to match those of the desired process.
As an example, the dependence of the decay ${\tilde X}^0_B\rightarrow \nu_{i}Z^0$, $Z^0\to\overline e e$--shown in Figure \ref{fig:BinoExample}--on $M_{{\tilde X}^0_B}$ may be extracted by calculating the dependence of the decay 
$\mu\to e \overline{\nu}_{e} \nu_\mu$ on the $\mu$ mass.
Up to differences in couplings, this is accomplished by setting $\hat m_{\overline{\nu}_e} = m_e$, $\hat\Gamma_{W^\pm}=\Gamma_{Z^0}$, $\hat M_{W^\pm}= M_{Z^0}$ in the calculation.
This method is used to calculate the partial widths for each of the Bino decays via $W^\pm$, $Z^0$, and $h^0$ bosons, taking into account masses for all products of the three-body decays.
\begin{figure}[t]
\centering
\begin{subfigure}[b]{0.8\textwidth}
\includegraphics[width=1.0\textwidth]{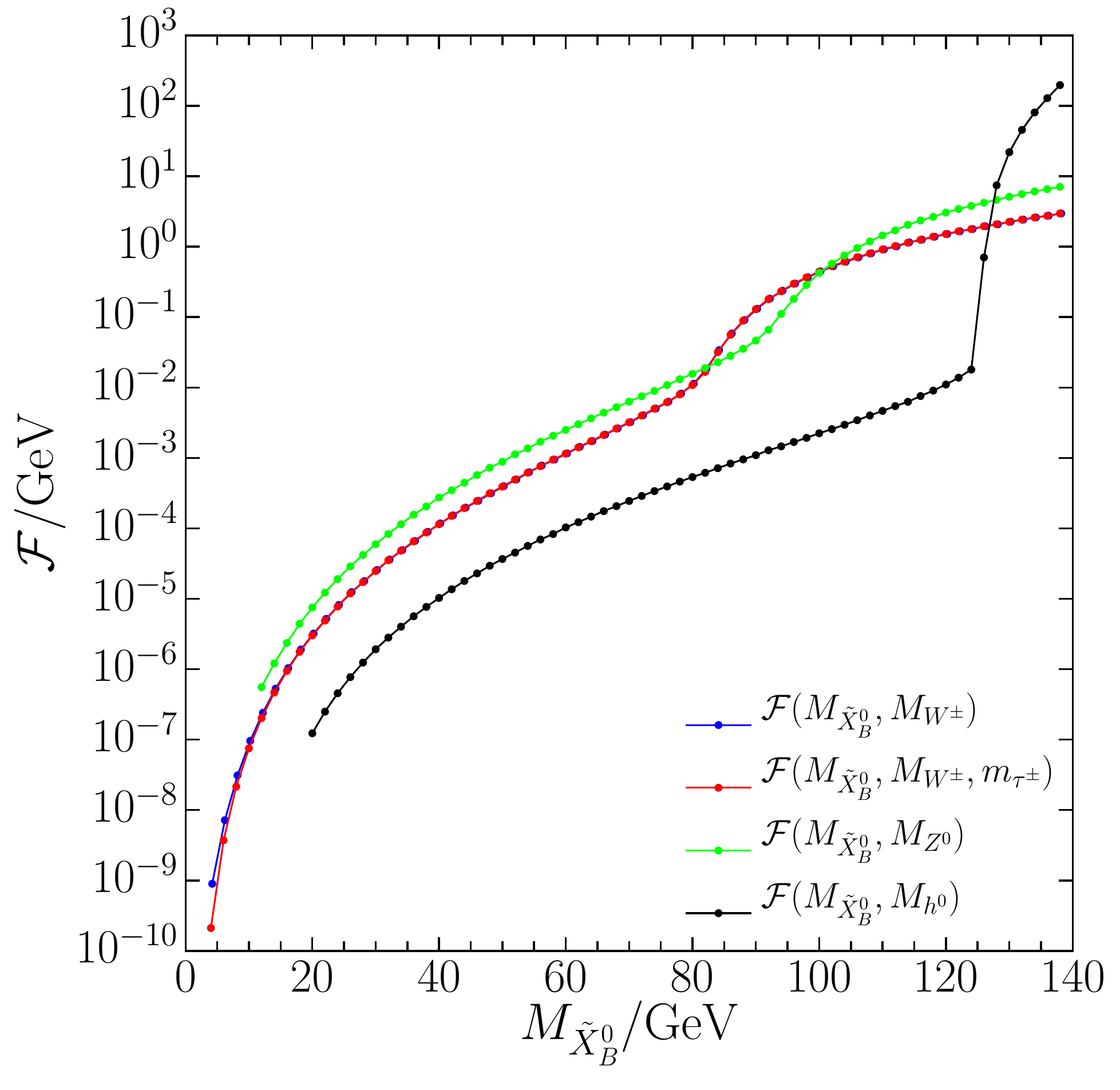}
\end{subfigure}
\caption{ Mass functions $\mathcal{F}(M_{\tilde X^0_{B}})$ defined analagously to \eqref{eq:def_f}, after summing over all final states from the electroweak boson decays.  Each function shows the dependence of the corresponding partial decay width on the mass of the decaying Bino LSP.  A single function is shown for each of the $Z^0$ and $h^0$ bosons plus neutrino decays, as all neutrino species have negligible mass. For Bino LSP decays to $W^\pm$ and a charged lepton, separate functions are shown for  ${\tilde X}^0_B\xrightarrow{W^\pm} e^\mp, \mu^{\pm}$ and ${\tilde X}^0_B\xrightarrow{W^\pm} \tau^\mp$ decays. While electron and muon masses are negligible, minor tau mass effects are visible in the case of very small Bino LSP masses. Note the rapid increase of each function as the LSP mass approaches, and then surpasses, the mass of the associated electroweak boson. For higher LSP mass, the decay can now proceed via an intermediate on-shell boson. For example, $d{\mathcal{F}}(M_{\tilde X^0_{B}}, M_{h^0})/dM_{\tilde X^0_{B}}$ increases rapidly near 125 GeV (the mass of the $h^0$ boson). Note that it is a more dramatic effect than in the case of the $W^\pm$ and $Z^0$ bosons, due to the fact that $\Gamma_{h^0} \ll \Gamma_{W^\pm}, \Gamma_{Z^0}$.}
\label{fig:fms}
\end{figure}

Returning to the decay rate \eqref{eq:Gamma_form}, we define
\begin{equation}\label{eq:def_f}
{\mathcal{ F}}(M_{\tilde X^0_B}, M_{Z^0}, m_{\nu_i})=\frac{1}{g_2^2}\sum_{f}g_f^2 F(M_{\tilde X^0_B}, M_{Z^0}, m_f, m_{\nu_i}),
\end{equation}
where the sum is over all possible decays of the $Z^0$ to fermion-antifermion pairs $\overline f f$. We choose to normalize $\mathcal F$ by dividing by $g_2^2$, where $g_2$ is the $SU(2)_L$ gauge coupling. 
Defining the process ${\tilde X}^0_B\xrightarrow{Z^0} \nu_{i}$ to be ${\tilde X}^0_B\xrightarrow{Z^0} \nu_{i}\overline f f$ summed over the final states $\bar{f}$ and $f$, it follows that the associated decay rate is given by
\begin{equation}\label{eq:Gamma_form2}
\Gamma_{{\tilde X}^0_B  \xrightarrow{Z^0} \nu_{i}}=g_{{\tilde X}^0_B\rightarrow Z^0 \nu_{i}}^2 g_{2}^{2}\mathcal{F}(M_{\tilde X^0_B}, M_{Z^0}, m_{\nu_i}).
\end{equation}
Similar definitions apply to decays which involve off-shell $W^\pm$ and $h^0$ bosons. 
Because the neutrino masses are negligible, the mass functions for ${\tilde X}^0_B\xrightarrow{Z^0} \nu_{i}$ and ${\tilde X}^0_B\xrightarrow{h^0} \nu_{i}$ decays are independent of the lepton family. Hence, we will write the associated ${\mathcal{F}}$ functions simply as ${\mathcal{ F}}(M_{\tilde X^0_B}, M_{Z^0})$ and ${\mathcal{ F}}(M_{\tilde X^0_B}, M_{h^0})$ 
respectively. For the decays ${\tilde X}^0_B\xrightarrow{W^{\pm}} \ell_{i}^{\mp}$, it is sufficient to treat both the electron and muon as massless and, hence, denote their ${\mathcal{F}} $ functions as ${\mathcal{ F}}(M_{\tilde X^0_B}, M_{W^{\pm}})$. However, the tauon mass cannot be neglected. Hence, a separate calculation is performed for ${\tilde X}^0_B\xrightarrow{W^\pm} \tau^\mp$. We will denote the associated ${\mathcal{F}}$ function as ${\mathcal{ F}}(M_{\tilde X^0_B}, M_{W^{\pm}}, m_{\tau^\pm})$.
Figure~\ref{fig:fms} shows the mass functions for all Bino decays, which are calculated using \madgraph\ and were defined analagously to \eqref{eq:def_f}.

\subsection{Lifetime of a light Bino LSP}

\begin{figure}[t]
   \centering

   \begin{subfigure}[c]{0.49\textwidth}
\includegraphics[width=1.0\textwidth]{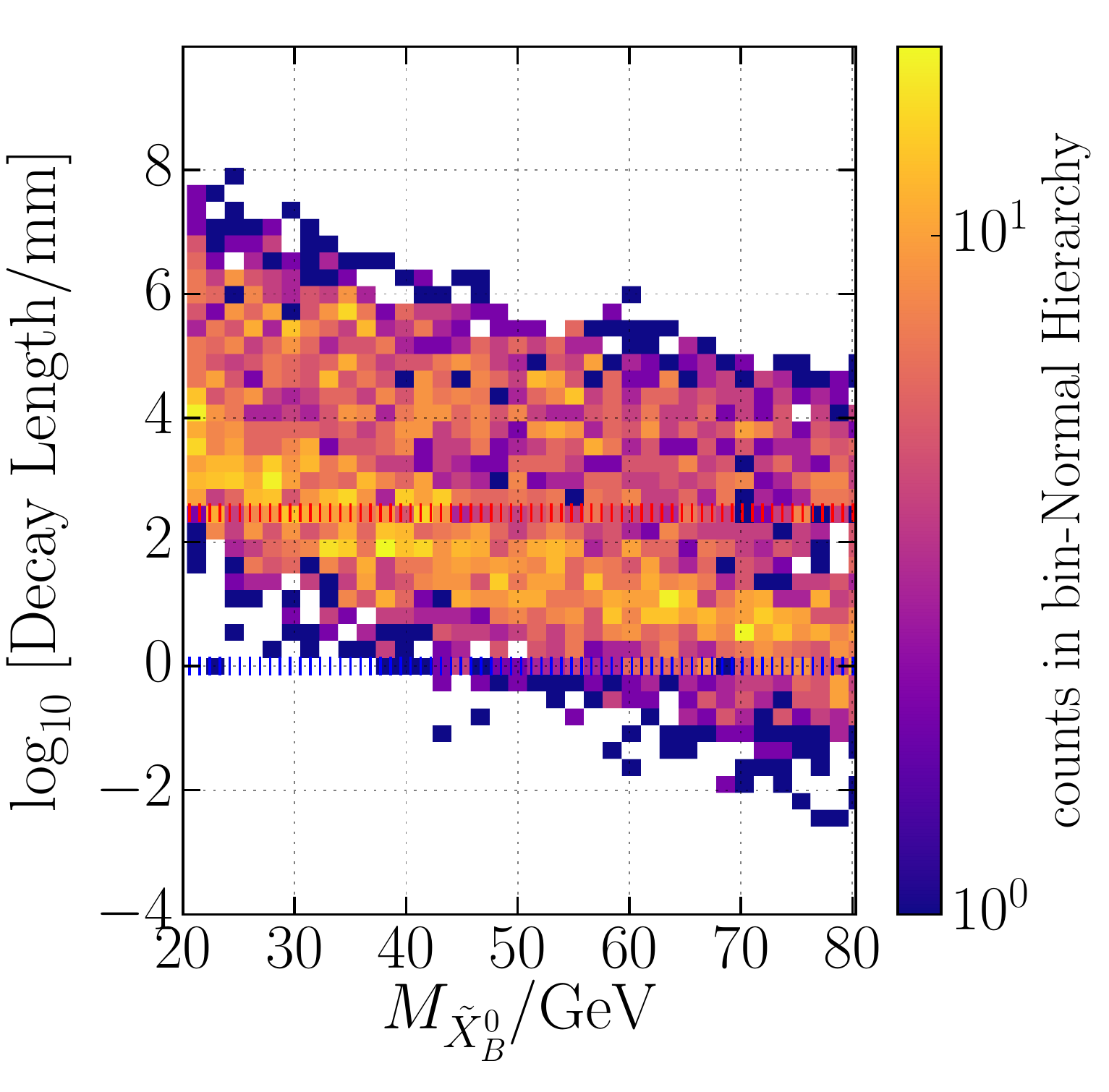}
\caption*{}
\label{fig:prompt_small_1}
\end{subfigure}
   \begin{subfigure}[c]{0.49\textwidth}
\includegraphics[width=1.0\textwidth]{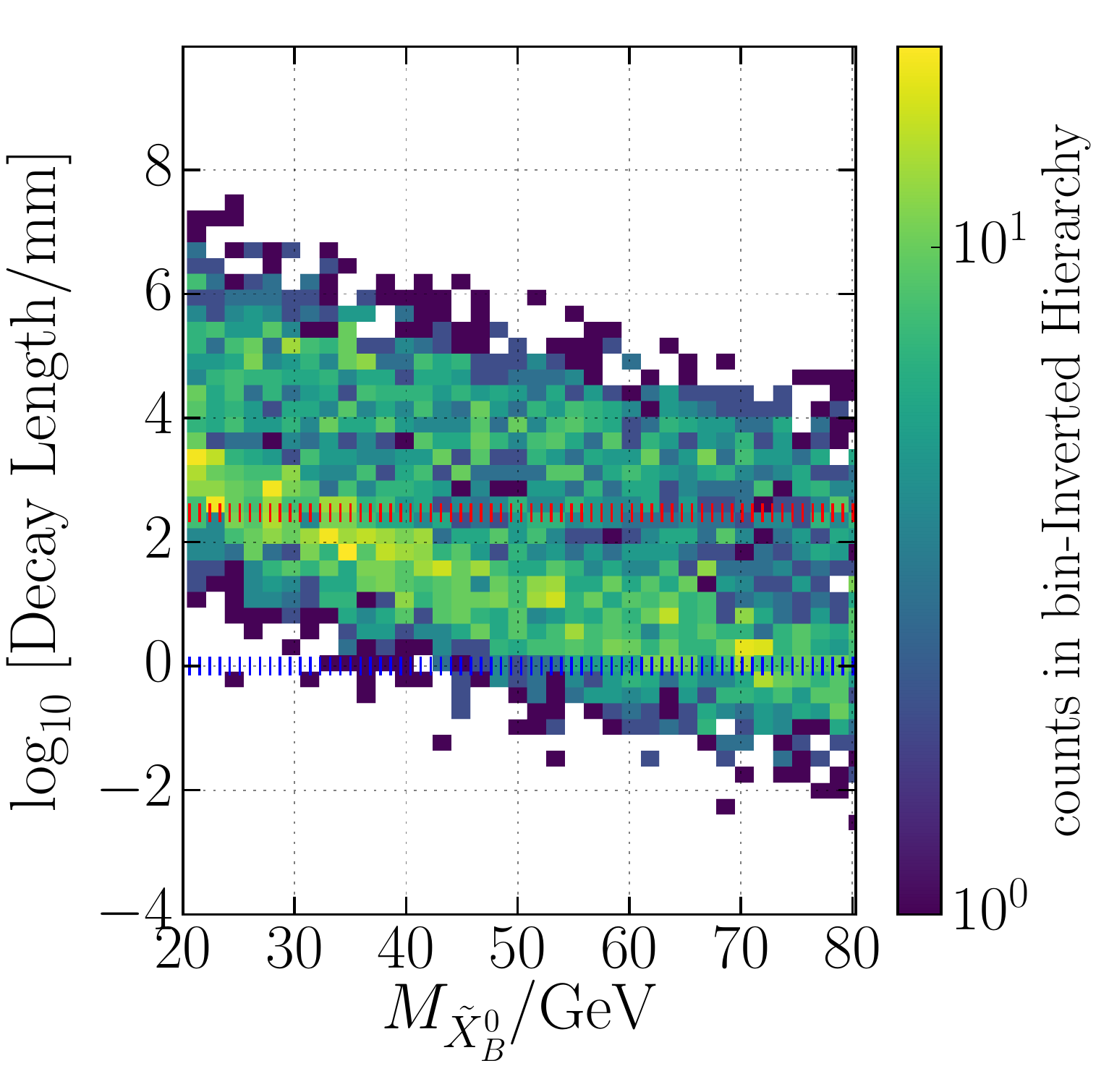}
\caption*{}
\label{fig:prompt_small_2}
\end{subfigure}
   \caption{Bino neutralino LSP RPV decay lengths, summed over all three channels, for Bino masses lighter than the $W^\pm$ and, hence, which can only decay through an off-shell boson. The results are in millimeters, for the normal and inverted hierarchies. The average decay length $L=c\times \frac{1}{\Gamma}$ increases for  smaller values of $M_{\tilde X^0_B}$. We have chosen $\theta_{23}=0.597$ for the normal neutrino hierarchy and $\theta_{23}=0.529$ for the inverted hierarchy to display the results. At and below the blue dashed line ($c\tau=1$~mm), the decays are considered prompt. 
   The red dashed line ($c\tau=30$~cm) denotes the largest decay lengths that may be measured via displaced vertices.}
   \label{fig:prompt_small_mass_total}
\end{figure}

In this section, we study whether light Bino neutralino LSPs can RPV decay promptly since their decays proceed only through {\it off-shell} bosons. We will compute these decay rates, summing over the partial widths of all possible final states produced. We use the mass functions $\mathcal{F}$ shown in Figure \ref{fig:fms}. The decay width of the Bino LSPs lighter than the EW scale is 
\begin{multline}
\Gamma_{{\tilde X}^0_B}=\sum_i \sum_{f} \Gamma_{{\tilde X}^0_B\xrightarrow{W^\pm} \ell^\mp_{i} \overline f^' f}+\sum_i \sum_{f} \Gamma_{{\tilde X}^0_B\xrightarrow{Z^0} \nu_{i} \overline f f}+\sum_i \sum_{f} \Gamma_{{\tilde X}^0_B\xrightarrow{h^0} \nu_{i} \overline f f}\\
=\sum_i g_{{\tilde X}^0_B\rightarrow W^\pm \ell^\mp_{i}}^2g_2^2 \mathcal{F}(M_{\tilde X^0_B}, M_{W^\pm}, m_{\ell_i^\pm})
+\sum_i g_{{\tilde X}^0_B\rightarrow Z^0 \nu_{i}}^2g_2^2 \mathcal{F}(M_{\tilde X^0_B}, M_{Z^0})+ \\ \sum_i g_{{\tilde X}^0_B\rightarrow h^0 \nu_{i}}^2g_2^2 \mathcal{F}(M_{\tilde X^0_B}, M_{h^0}).
\end{multline}
The decay lengths of the Bino LSPs,
\begin{equation}
L=\frac{c}{\Gamma_{{\tilde X}^0_B}},
\end{equation}
are calculated for both the normal and inverted hierarchy scenarios and shown in Figure \ref{fig:prompt_small_mass_total}. Prompt decays ($L<1$mm) are possible for Bino neutralino LSP masses as low as about 50~GeV, in both the normal and the inverted hierarchy scenarios. However, such Bino LSPs are most likely to decay with significant displacement from the production vertex, though still within the typical LHC detector volume. Bino LSPs with masses in the range from about 50~GeV to 20~GeV do not exhibit prompt decays, but can still decay with displaced vertices in the detector. 
Bino LSPs with very low masses (roughly $<20$~GeV) may be stable on the scale of LHC detectors.
Conventional ``missing-energy'' searches should have some sensitivity to these models, 
while ambitious next-generation experiments~\cite{Chou:2016lxi} may offer the possibility for direct detection of displaced decays. 

Note that Figure \ref{fig:prompt_small_mass_total} displays the decay lengths strictly for Bino LSPs lighter than $M_W^\pm$, which can only decay via off-shell processes. However, decays via on-shell $Z^0$ bosons become forbidden even earlier; that is, for Bino LSPs lighter than $M_{Z^0}$.  Similarly, Bino LSPs with masses smaller than $M_{h^0}$ cannot decay through an on-shell Higgs bosons. Therefore, Bino LSPs with masses in the interval between $M_{W^{\pm}}$ and $M_{h^0}$ could possibly, for example, decay via both on-shell $W^{\pm}$ bosons and off-shell $Z^0, \> h^0$ bosons. However, in this region, decays via off-shell $Z^0, \> h^0$ bosons are strongly suppressed in general compared to the decays via the on-shell $W^{\pm}$ bosons. 
The effect of this suppression is seen in Figure \ref{fig:fms}. For decays via the $W^\pm$ boson, the red curve drops about two orders of magnitude when we move from the on-shell region, where the mass of the incoming Bino is larger than $M_{W^\pm}$, to the off-shell one,  where the mass of the incoming Bino is smaller than $M_{W^\pm}$.  A similar drop in magnitude occurs in the green line for decays via an on-shell versus an off-shell $Z^0$ boson. Even more pronounced is the drop from the on-shell to the off-shell region, approximately four orders of magnitude, for the decays via a Higgs boson-- the black curve in Figure \ref{fig:fms}.  It follows that for a Bino LSP mass above $M_{W^\pm}$, but below $M_{Z^{0}}$ and  $M_{h^0}$, the size of the $\mathcal{F}$ functions for $Z^{0}$, $h^{0}$ are significantly suppressed relative to $\mathcal{F}$ for the $W^{\pm}$. Hence, in this mass regime, the decay rate of the Bino LSP is dominated by decay via an on-shell $W^{\pm}$; the decay rates for the off-shell $Z^{0}$ and, particularly, the off-shell $h^{0}$ being suppressed.
Note that in all three cases the {\it transition interval} from on-shell to off-shell bosons is narrow, of order $\approx 10$ GeV.

Considering the relatively narrow transitions between the on-shell to the off-shell regions for all decay channels and the strong suppression of the off-shell processes, we neglect the off-shell decays via the $Z^0$ and $h^0$ bosons when decays via on-shell $W^\pm$ bosons are possible. Figure \ref{fig:LSPprompt_neut}, which takes into account only processes that occur via on-shell bosons, provides accurate estimates for the summed decay lengths of all Binos heavier than $M_{W^{\pm}}$. Figure \ref{fig:prompt_small_mass_total}, which presents the summed decay lengths for all Bino LSPs lighter than $M_{W^\pm}$,  completes the decay width analysis.

\begin{figure}[t]
\centering
\begin{subfigure}[b]{1.\textwidth}
\includegraphics[width=1.0\textwidth]{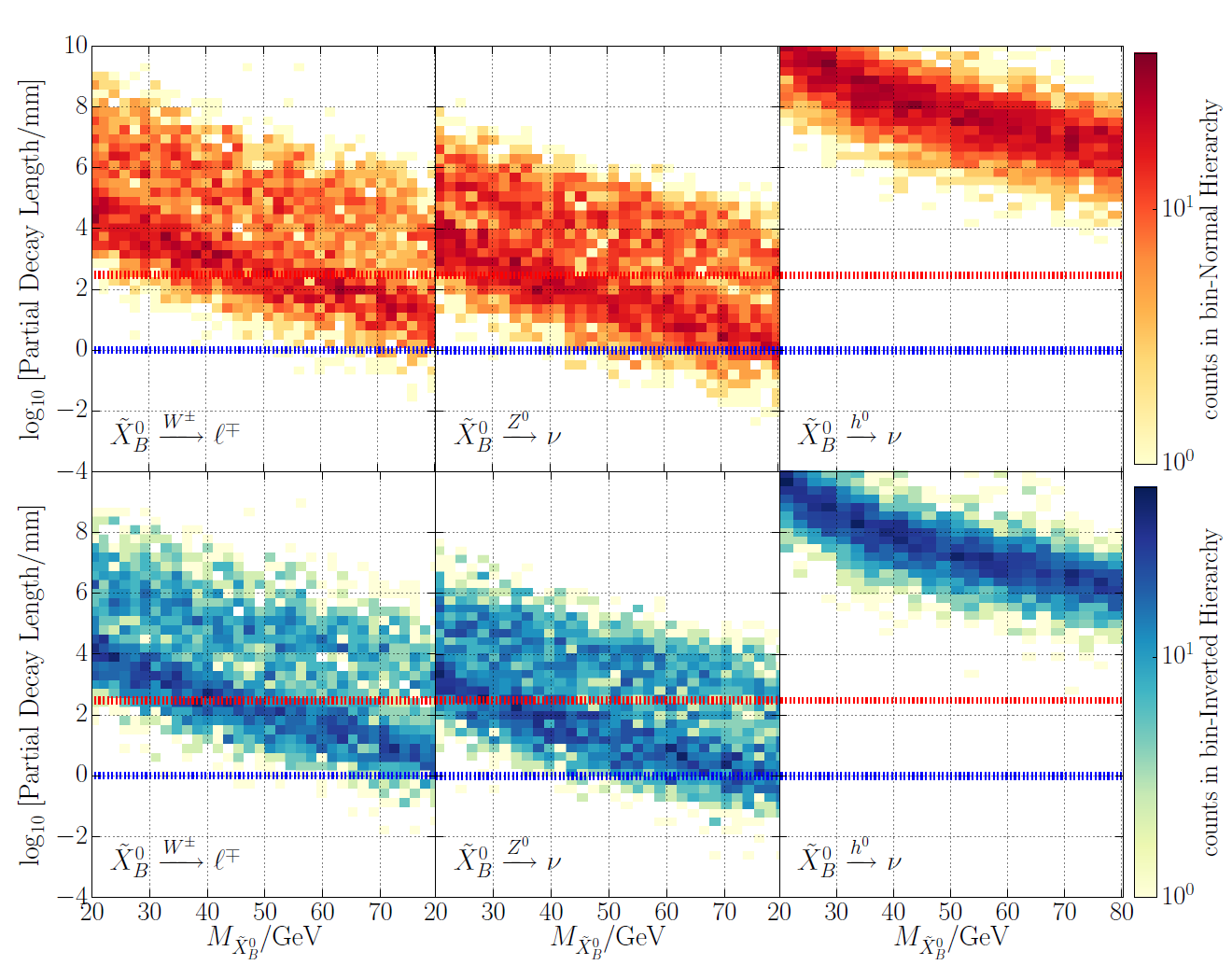}
\end{subfigure}
\caption{ Bino neutralino LSP partial decay lengths in millimeters, for individual decay channels, for both normal and inverted hierarchies.Widths are calculated for Bino masses when all decays must proceed through intermediate off-shell bosons. We have chosen $\theta_{23}=0.597$ for the normal neutrino hierarchy and $\theta_{23}=0.529$ for the inverted hierarchy. At and below the blue dashed line ($c\tau=1$~mm), the decays are considered prompt. The red dashed line ($c\tau=30$~cm) denotes the largest decay lengths that may be measured via displaced vertices.}
\label{fig:PromptnessSmall}
\end{figure}

\subsection{Branching ratios of the Bino LSP RPV decays}

For Bino LSPs masses smaller that the mass of $W^{\pm}$, there is a wide range that could lead to visible signatures in LHC detectors-- despite decaying via off-shell bosons.  Therefore, we separately analyze each of the decay channels, to determine the dominant decay modes. To mimic the analysis undertaken in Section 2, we classify the Bino neutralino decays into three categories, depending on which off-shell boson, $W^\pm, \> Z^0$ or $h^0$, the Bino neutralino LSP decays into. For each category, we compute the decay rates by summing over the three lepton families produced in the Bino decay and over all final-state particles associated with the electroweak boson decay.
For example, the partial decay length $L$ of a Bino LSP associated with decays through an off-shell $Z^0$ boson is 
\begin{equation}
L_{\tilde X_B^0 \xrightarrow{Z^0} \nu}=\frac{c}{\Gamma_{\tilde X_B^0 \xrightarrow{Z^0} \nu}},
\end{equation}
where
\begin{equation}
\Gamma_{{\tilde X}^0_B\xrightarrow{Z^0} \nu}=\sum_{i} \sum_{f} \Gamma_{{\tilde X}^0_B\xrightarrow{Z^0} \nu_{i} \bar{f} f} \ .
\end{equation}
\begin{figure}[t]
\centering
\begin{subfigure}[b]{1.\textwidth}
\includegraphics[width=1.0\textwidth]{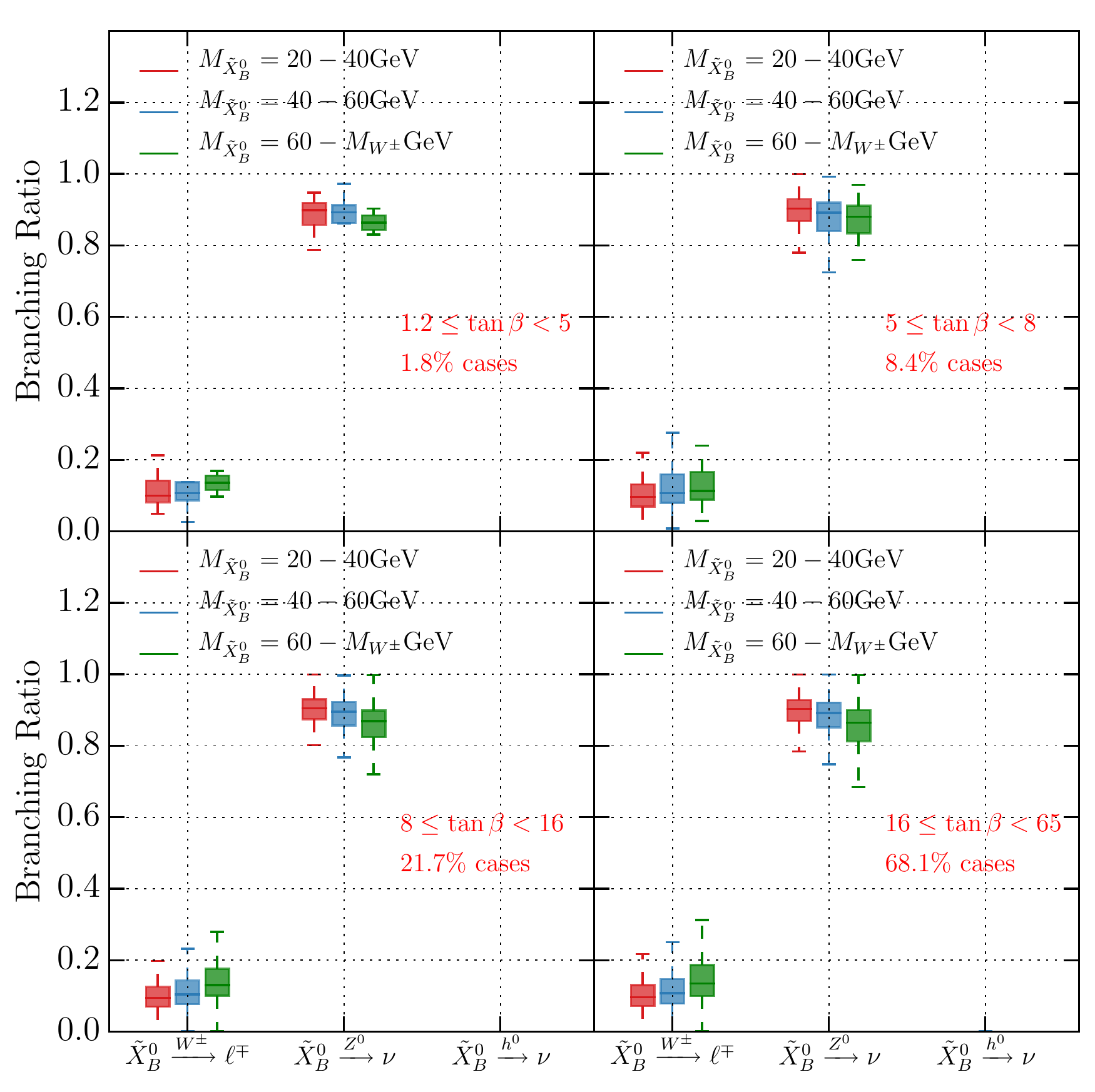}
\end{subfigure}
\caption{ Branching ratios for the three possible decay channels of a Bino neutralino LSP divided over three mass bins and four $\tan \beta$ regions. We studied the viable points for Bino LSPs with masses smaller than the mass of the $W^\pm$ bosons but larger than 20~GeV. The colored horizontal lines inside the boxes indicate the median values of the branching fraction in each bin, the boxes indicate the interquartile range, while the dashed error bars show the range between the maximum and the minimum values of the branching fractions. The case percentage indicate what percentage of the physical mass spectra have $\tan \beta$ values within the range indicated.  We assumed a normal neutrino hierarchy, with $\theta_{23}=0.597$. Note that the branching ratios via an off-shell $h^{0}$, while non-vanishing, are of order $10^{-3}$ and smaller and, hence, too small to appear in the Figure. }\label{fig:BarSmall}
\end{figure}
The results are shown in Figure \ref{fig:PromptnessSmall}. We learn that Bino LSPs lighter than $M_{W^\pm}$ decay mainly via $W^\pm$ and $Z^0$ bosons, with decays proceeding via off-shell Higgs being completely negligible. It is also {\it important to note} that for Bino masses below approximately 20~GeV, all three channels have decay lengths larger than 30 cm--that is, are longer than displaced vertices--and, hence, are essentially stable within the ATLAS detector. This is consistent with our previous observation from Figure \ref{fig:prompt_small_mass_total}.

In Figure \ref{fig:BarSmall}, we analyze the branching ratios of the three main decay channels of these light Bino LSPs. We group the light Bino LSPs into three mass bins and four $\tan \beta$ bins, as we did for the heavier Bino LSPs in Figure \ref{fig:bar_plot2}. For consistency, we keep the same binning for the $\tan \beta$ parameter, and choose evenly spaced mass bins
\begin{equation}
[20\>\text{GeV}, 40\>\text{GeV}], \quad[40\>\text{GeV}, 60\>\text{GeV}], \quad[\text{60}\>\text{GeV}, M_{W^\pm}]. 
\end{equation}
Note that, as discussed above, the decay lengths for Bino LSP masses below 20 GeV are all generally very large and outside the detector. We therefore don't consider Bino masses smaller than 20~GeV in our analysis.
 We learn that Bino LSPs decay mainly via an off-shell $Z^0$ boson independent of mass. This type of decay is usually at least four times more probable than decays via off-shell $W^\pm$ bosons. Decays via off-shell Higgs, although non-vanishing, are significantly suppressed and thus not likely to be observed.
Variations in $\tan\beta$ do not significantly affect the Bino LSP branching fractions.

\subsection{Experimental signatures of off-shell Bino LSP decays}
So far, we have seen that Bino LSPs as light as 20 GeV may be detected at the LHC if they decay via $W^\pm$ or $Z^0$ bosons. In this subsection, we analyze the experimental signature of such decays in the detector. That is, we compute two sets of branching ratios: one corresponding to the family of the lepton produced at the RPV vertex and another for the decay products of the electroweak boson. 

Note that our analysis of the Bino LSPs decays when they are lighter than the electroweak vector bosons differs slightly from our approach in Section 4, where we studied the RPV decays of the Bino LSPs heavier than the electroweak scale. In Section 4, we only considered the vertices shown in Figure \ref{fig:NeutralinoDecays}, in which the Bino decays into an on-shell boson and a lepton via RPV couplings. The physical boson that is produced would further decay into pairs of final states $\bar{f}$ and $f$. However, we did not discuss such decays, since these standard model processes are well-known. Hence, we limited ourselves to analyzing the statistical distributions for the families of the leptons produced at the RPV vertex only.

However, for decays via off-shell bosons, we need to consider the full diagrams shown in Figure \ref{fig:BinoDecays2}; that is, including the RPV vertex, the off-shell boson propagator and the vertex in which the final states $\bar{f}$ and $f$ are produced. 
Equation \eqref{eq:Gamma_form} shows that the decay rate for each such individual process is proportional to the RPV coupling, the $g_{f}$ coupling and a function $F$ which depends on the masses of the initial, intermediate and final states. The final states  $\bar{f}$ and $f$ at the second vertex and the lepton produced at the first vertex are interconnected through the function $F$, obtained after integrating over the momenta of all the final and initial states. An unified treatment for the branching fractions at the first and second vertices would be very complicated, as it would involve counting all possible combinations of decay products at these vertices. However, in the physical regime that we are working in, that is, for incoming Binos heavier than 20 GeV,  a simplifying assumption can be made. Because the final decay products are much lighter than the incoming Bino, the function $F$ has a weak dependence on the masses of these final states. Therefore, the two sets of branching ratios depend mainly on the value of the coupling at each vertex and become independent of each other. We will now explain how accurate this assumption is and what sets of branching fractions we expect at each of the two vertices.

First, we analyze the distributions of the branching ratios to different lepton families at the RPV vertex. In Figure \ref{fig:neutralino_lepton_family}, we have shown the results of a similar study, but in the case of heavy Bino LSPs  that could decay via on-shell bosons. When repeating the computation for light Bino LSPs with masses in the detectable interval from 20~GeV to $M_{W^\pm}$, we obtain statistically identical distributions to those shown in Figure \ref{fig:neutralino_lepton_family}. The similarity between the distributions in the off-shell and the on-shell regime is expected. The reason is that when calculating the branching fractions at the RPV vertex in the off-shell decay regime for a fixed pair of final states at the second vertex,  the $F$ functions and the couplings at the second vertex are divided out.   The cancellation is possible because the $F$ functions show no dependence on the family of the lepton produced at the first vertex. Therefore, the  generation of leptons produced at the first vertex depends on the RPV breaking parameters  $\epsilon_i$ and $v_{L_i}$ only,  just as in the on-shell decay regime. This result is independent of the nature of the particles $f$ produced at the second vertex.  Furthermore, the relative sizes of the RPV breaking parameters do not depend on the Bino mass and, therefore, the branching fraction distributions at the first vertex are identical to those in Figure \ref{fig:neutralino_lepton_family}.

Secondly, we compute the expected branching ratios to different pairs of final states  $\bar{f}$ and $f$ at the second vertex. This time, we fix the family of the lepton produced at the first vertex. The RPV couplings from the first vertex are divided out. The $F$ functions do not depend on the family of the lepton produced at the first vertex. Hence, the set of branching ratios to final states  $\bar{f}$ and $f$ at the second vertex is independent of the nature of the lepton at the first vertex. Note that when we computed the branching ratios at the first vertex for a fixed pair of final states  $\bar{f}$ and $f$, we divided out the $F$ functions as well, because they showed no dependence on the lepton family. However, the $F$ functions depend on the heavier pair of states  $\bar{f}$ and $f$, although weakly. If this was not the case, we would simply recover the same branching fractions at the second vertex as those calculated for physical $W^\pm$, $Z^0$ and $h^0$ bosons, existent in the standard model literature. Although the effect is weak, these branching fractions at the second vertex differ from their on-shell values, especially for the lightest Bino LSPs. As the mass $M_{\tilde X_B^0}$ of the incoming Bino is taken to be lighter, it becomes comparable to some of the masses of the $W^\pm$ and $Z^0$ decay products. Hence, the decays to these final states  $\bar{f}$ and $f$ get more and more suppressed.
To illustrate this effect, in Tables \ref{table:Zboson} and \ref{table:Hboson} we compare the relative fractions for each of the electroweak boson decays, for Bino LSP masses of 30 GeV and 60 GeV .  For comparison, in the columns labelled $M_{\tilde X_B^0}>M_{Z^0}$  and $M_{\tilde X_B^0}>M_{h^0}$ of Tables \ref{table:Zboson} and \ref{table:Hboson} respectively, we show the SM branching fractions calculated for on-shell $Z^0$ and $h^0$.

\begin{table}[t]

\begin{center}
  \begin{tabular}{ l |c|c|c }
    \hline
    \multirow{2}{*}{Process} &\multicolumn{3}{c}{$Z^0$ branching fractions for $Z^0 \rightarrow \overline{f} f$ decays (in \%)} \\
    \cline{2-4}
     & $M_{\tilde X_B^0}>M_{Z^0}$ & $M_{\tilde X_B^0}=60$~GeV & $M_{\tilde X_B^0}=30$~GeV \\ 
    \hline
    $Z^0 \rightarrow e^\pm e^\mp $       & \phantom{0}3.4 & \phantom{0}3.4 & \phantom{0}3.6  \\
    $Z^0 \rightarrow \mu^\pm \mu^\mp$    & \phantom{0}3.4 & \phantom{0}3.4 & \phantom{0}3.6  \\
    $Z^0 \rightarrow \tau^\pm \tau^\mp $ & \phantom{0}3.4 & \phantom{0}3.4 & \phantom{0}3.4  \\
    \hline %
    $Z^0 \rightarrow \nu \overline \nu $ & 20.0 &  20.3 &  21.3  \\
    \hline %
    $Z^0 \rightarrow u \overline u $     & 11.6 &  11.7 &  12.4  \\
    $Z^0 \rightarrow c \overline c $     & 12.0 &  12.0 &  12.2  \\
    \hline %
    $Z^0 \rightarrow d \overline d $     & 15.6 &  15.8 &  16.6  \\
    $Z^0 \rightarrow s \overline s $     & 15.6 &  15.8 &  16.6  \\
    $Z^0 \rightarrow b \overline b $     & 15.1 &  14.3 &  10.3  \\
    \hline
  \end{tabular}
\end{center}
\caption{Branching fractions for decays of the virtual $Z^0$ boson for the $Z^0 \rightarrow \overline{f} f$  process for several values of the $\tilde X_B^0$ mass. The reference values for on-shell $Z^0$ decays ($M_{\tilde X_B^0}>M_{Z^0}$) are taken from the Particle Data Group recommendations~\cite{PDG}.}
\label{table:Zboson}
\end{table}

\begin{table}[t]
\begin{center}
  \begin{tabular}{ l |c|c|c }
    \hline
    \multirow{2}{*}{Process} &\multicolumn{3}{c}{$h^0$ branching fractions for $h^0 \rightarrow \overline{f} f$  decays (in \%)} \\
    \cline{2-4}
     & $M_{\tilde X_B^0}>M_{h^0}$ & $M_{\tilde X_B^0}=60$~GeV & $M_{\tilde X_B^0}=30$~GeV \\ 
    \hline %
    $h^0 \rightarrow b\bar b $          & \phantom{}58.9   & \phantom{}84.2   & \phantom{}87.0   \\
    $h^0 \rightarrow c\bar c $          & \phantom{0}2.9   & \phantom{0}4.1   & \phantom{0}4.4   \\
    $h^0 \rightarrow \tau^\pm \tau^\mp$ & \phantom{0}6.3   & \phantom{0}8.0   & \phantom{0}6.9   \\
    $h^0 \rightarrow \mu^\pm \mu^\mp$   & \phantom{00}0.02 & \phantom{00}0.03 & \phantom{00}0.02 \\
    \hline %
    $h^0 \rightarrow gg $               &  \phantom{0}7.8   & \phantom{0}3.5    & \phantom{0}1.7    \\
    $h^0 \rightarrow W^\pm W^\mp$       &  \phantom{}21.0   & \phantom{00}0.02  & \phantom{}$<0.01$ \\
    $h^0 \rightarrow Z^0 Z^0$           &  \phantom{0}2.6   & \phantom{00}0.01  & \phantom{}$<0.01$ \\
    $h^0 \rightarrow \gamma\gamma$      &  \phantom{00}0.23 & \phantom{00}0.07  & \phantom{00}0.01  \\
    $h^0 \rightarrow Z^0\gamma$         &  \phantom{00}0.15 & \phantom{}$<0.01$ & \phantom{}$<0.01$ \\
   \hline
  \end{tabular}
\end{center}
\caption{Branching fractions for decays of the virtual Higgs boson for the $h^0 \rightarrow \overline{f} f$  process for several values of the $\tilde X_B^0$ mass. These values are adapted from the Higgs branching fractions presented as a function of mass, published by the CERN LHC Higgs Yellow Report~\cite{deFlorian:2016spz}. Decay modes which contribute $<0.01$\% for all values of the Bino LSP mass  are suppressed.}
\label{table:Hboson}

\end{table}

For Bino LSP decays via $Z^0$ bosons, shown in Table~\ref{table:Zboson}, the partial widths corresponding to quark-antiquark pairs are somewhat suppressed relative to the SM decays. This is most apparent in the decays to $b\bar b$, but this suppression also occurs in the widths corresponding to charm and tau decays, but to a lesser degree.  For decays via $W^\pm$ bosons, all final state particles are light enough that the impact of the Bino LSP mass on the relative branching fractions is negligible for the considered mass range, so a dedicated Table is not presented.
Bino LSP decays via the Higgs boson are very rare, as discussed above.
However, the relative branching fractions of these processes are given in Table~\ref{table:Hboson}, to compare the importance of each channel to the final result.  These figures are adapted from Higgs decay widths calculated in the CERN Yellow Report~\cite{deFlorian:2016spz} for various values of the Higgs boson mass.

\subsection{Experimental outlook}

These findings demonstrate that the Bino LSP is a viable candidate for direct detection at the LHC across a wide range of masses.
For very low masses of the Bino LSP, the existing search program for $R$-parity conserving SUSY scenarios should be sensitive to final states with this new ``detector-stable'' particle.
Such searches may also be sensitive in the case of prompt Bino LSP decays to neutrinos (that is,  $\tilde X_B^0 \rightarrow Z^0 \nu$ and $\tilde X_B^0 \rightarrow h^0 \nu$).
Generally, a diverse set of searches for $R$-parity violating decays using prompt objects should also be pursued. 
In particular, maximal sensitivity could be obtained by taking advantage of the unconventional signatures produced in Bino LSP decays, such as $W^\pm$-lepton resonances.
Finally, the calculated distribution of possible lifetimes makes it abundantly clear that searches for displaced leptons and jets are an invaluable tool, particularly when the Bino LSP is lighter than the $W^\pm$ boson.

The Bino presents an attractive candidate to the experimentalist, as it is by far the most prevalent LSP in the space of models considered in the present analysis.
As has been shown, it may also be arbitrarily lighter than the soft SUSY breaking scale, due to cancelling contributions from unrelated soft mass terms.
On the other hand, pure Bino pairs cannot be produced directly from SM particle decays, so that experimental prospects will in general depend on the detailed spectrum of heavier SUSY particles.
However, this makes the prediction of a long-lived Bino LSP intriguing, as it is a process with no SM background.
This enables Bino LSP searches to be conducted without regard to the potentially complicated mechanism responsible for their production.
Hence, searches for displaced leptons and jets (independent of other activity in the detector) present a completely orthogonal method of probing otherwise challenging spectra of sparticle masses.

\section{Conclusion}

In this paper, using the formalism developed in \cite{Ovrut:2015uea,Dumitru:2018jyb}, we have shown that the Bino neutralino is the most prevalent LSP of the $B-L$ MSSM. An accurate approximation to its mass formula is presented and compared to the mass formula for both Wino charginos and Wino neutralinos, that were discussed in detail in a previous paper \cite{Dumitru:2018nct}. It is shown that, whereas the Wino LSP masses must always exceed the $W^{\pm}$ electroweak boson mass, the mass of the Bino neutralino LSP, while generically also larger than $M_{W^{\pm}}$, can be smaller than this scale--although such ``light'' Binos  are less prevalent. The mass spectrum for the Bino neutralino LSP is displayed. We have shown, however, that for sufficient ``fine-tuning'' its mass can actually become vanishly small. 

We then proceed to analyze the decays channels, decay rates/lengths and branching ratios for the RPV decays of Bino neutralino LSPs in the $B-L$ MSSM. This analysis, following the above comments, naturally breaks into  two different parts: a) for the Bino neutralino mass $M_{{\tilde{X}}_{B}^{0}} > M_{W^{\pm}}$ and b) for  $M_{{\tilde{X}}_{B}^{0}} < M_{W^{\pm}}$. Since the Bino neutralino mass can be made arbitrarily small by fine-tuning, in this paper we put a lower bound of 20 GeV on its mass for two reasons--1) since below that value the degree of fine-tuning increases dramatically and 2) when $M_{{\tilde{X}}_{B}^{0}} < 20$~GeV its decay length becomes very large, outside the range of the ATLAS detector. The mass of the Bino neutralino LSP has an important impact on its RPV decays. For $M_{{\tilde{X}}_{B}^{0}} > M_{W^{\pm}}$, it can always directly decay to a lepton and at least one, and perhaps each, of the three {\it on-shell} $W^{\pm}$, $Z^{0}$and $h^{0}$ bosons. In this regime, we compute the branching ratios for each boson decay channel. The associated decay lengths are also presented, both summing over all three decay channels and for each channel independently. A discussion of whether the decays are ``prompt'', occur as ``displaced vertices'' or are longer is given. We also analyze the branching fractions for each boson channel into individual leptons. Finally, the relationship of the decay lengths and the individual branching fractions to the neutrino mass hierarchy--both normal and inverted, is discussed in detail.

For Bino neutralino LSPs with mass in the range $[20~{\rm GeV},M_{W^{\pm}}]$, the RPV decays must occur via one of three {\it off-shell} $W^{\pm}$, $Z^{0}$and $h^{0}$ bosons.
The analysis of decays channels, decay rates/lengths and branching ratios for these RPV off-shell processes is much more computationally involved. Our method of calculation is presented and used to compute the same quantities as in the on-shell case. The fact that the intermediate bosons are off-shell significantly lowers the decay rates--and, hence, there are fewer prompt decays in this category, most lengths being at least displaced vertices and much larger. However, the effect of the type of neutrino hierarchy does not greatly change from the previous analysis. The branching fractions to a specific lepton at the first RPV vertex is almost unchanged from the heavy Bino case. However, the analysis of the decay products arising from the decay of the off-shell boson does somewhat change. The branching fractions for these decays are analyzed separately.

We conclude that for an LSP Bino neutralino in the $B-L$ MSSM there is, regardless of its mass, a significant chance that its RPV decays through various specified channels can be observed in the run 2 data at the LHC. If discovered, the theoretical predictions presented here could be a first discovery of possible $N=1$ supersymmetry in nature and, secondly, partially validate the specific $B-L$ MSSM theory. 

\section*{Acknowledgments}
The authors would like to thank Evelyn Thomson, Elliot Lipeles, Jeff Dandoy, Christopher Mauger, Nuno Barros, Austin Purves and Leigh Schaefer for helpful suggestions. Ovrut would also like to acknowledge many informative conversations with Zachary Marshall and Sogee Spinner concerning RPV decays of a stop LSP. Burt Ovrut, Sebastian Dumitru are supported in part by DOE No. DE-SC0007901 and SAS Account 020-0188-2-010202-6603-0338. Christian Herwig is supported in part by DOE No. DE-SC0007901.

\begin{appendices}

\section {Notation}\label{appendix:notation}

In this Appendix, we present for clarity all the notation used throughout the paper.
\subsection{Gauge Eigenstates}

\begin{itemize}
\item{Bosons}

\underline{{\it vector gauge bosons}}

~~~ $SU(2)_L-\quad\> W^1_\mu\>, W^2_\mu\>, W^3_\mu$, $\quad $   
coupling parameter $g_2$

~~~ $U(1)_{B-L}-\quad \> B^'_\mu \> $, $\quad $   
coupling parameter $g_{BL}$

~~~ $U(1)_{3R}-\quad \> {W_R}_\mu \> $, $\quad $   
coupling parameter $g_R$

~~~ $U(1)_{Y}-\quad \> {B}_\mu \> $, $\quad $   
coupling parameter $g'$

~~~ $U(1)_{EM}-\quad \> {\gamma}^0_\mu \> $, $\quad $   
coupling parameter $e$

~~~B-L Breaking:    $U(1)_{3R}\otimes U(1)_{B-L}\rightarrow U(1)_Y,\quad$    massive boson ${Z_R}_\mu$,$\>\>$ coupling $g_{Z_R}$

~~~EW Breaking:    $SU(2)_L\otimes U(1)_Y \rightarrow U(1)_{EM},\quad$    massive bosons $Z^0_\mu
,\>W^\pm_\mu\quad$

\underline{{\it Higgs scalars}}

~~~$H_u^0\>, H_u^+\>, H_d^0\>, H_d^-\quad $

\item{Weyl Spinors}

\underline{ {\it gauginos}}

~~~ $SU(2)_L-\> \tilde W^0\>, \tilde W^\pm$,\quad
$U(1)_{B-L}- \>\tilde B^' , \quad$ $U(1)_{3R}-\> {\tilde W_R} \> $,\quad
$U(1)_{Y}-\> \tilde {B},\quad$ $U(1)_{EM}- \> \tilde {\gamma}^0 \> $

\vspace{1mm}
\underline{{\it Higgsinos}}

~~~$\tilde H_u^0\>, \tilde H_u^+\>, \tilde H_d^0\>, \tilde H_d^-$

\underline{{\it leptons}}

 ~~~left chiral\quad  $e_i,\>\nu_i, \>\> i=1,2,3 \quad \text{where} \quad e_1=e,\>e_2=\mu,\>e_3=\tau$
          
 ~~~right chiral\quad  $e^c_i,\>\nu^c_i, \>\> i=1,2,3 \quad \text{where} \quad e^c_1=e^c,\>e_2^c=\mu^c,\>e_3^c=\tau^c$

\underline{{\it sleptons}}

 ~~~left chiral\quad  $\tilde e_i,\>\tilde \nu_i, \>\> i=1,2,3 \quad \text{where} \quad \tilde e_1=\tilde e,\>\tilde e_2=\tilde \mu,\>\tilde e_3=\tilde \tau$
          
 ~~~right chiral\quad  $\tilde e^c_i,\>\tilde \nu^c_i, \>\> i=1,2,3 \quad \text{where} \quad \tilde e^c_1=\tilde e^c,\>\tilde e_2^c=\tilde \mu^c,\>\tilde e_3^c=\tilde \tau^c$

\end{itemize}

\subsection{Mass terms}

\indent \quad \quad \underline{ {\it gauginos}}

~~~ $\tilde W^0\>, \tilde W^\pm \rightarrow M_2,\quad$
$\tilde B^'\rightarrow M_{BL} , \quad$ $ {\tilde W_R} \rightarrow M_R,\quad $ $\tilde {B} \rightarrow M_1,\quad$ 

\vspace{1mm}

\underline{{\it Higgsinos}}

~~~$\tilde H_u^0\>, \tilde H_u^+\>, \tilde H_d^0\>, \tilde H_d^- \rightarrow \mu$

\vspace{1mm}

\underline{{\it left chiral charged leptons}}

~~~$e_i \rightarrow m_{e_i}, \>$ for $i=1,2,3$

\vspace{1mm}

\underline{{\it right chiral charged leptons}}

~~~$e_i^c \rightarrow m_{e_i^c}, \>$ for $i=1,2,3$

\subsection{Mass Eigenstates}

\begin{itemize}

\item{Weyl Spinors}

\underline{{\it leptons}}\\
 \quad  $e_i,\>\nu_i, \>\> i=1,2,3 \quad \text{where} \quad e_1=e,\>e_2=\mu,\>e_3=\tau$

\underline{{\it charginos and neutralinos}}\\
  \quad $\tilde \chi^\pm_1, \quad \tilde \chi^\pm_2, \quad \tilde \chi_n^0,\quad n=1,2,3,4,5,6$

\item{4-component Spinors}

\underline{{\it  leptons}}\\
~~~$\ell_i^-=\left(\begin{matrix}e_i\\ {e_i^c}^\dag\end{matrix}\right),\quad
\ell_i^+=\left(\begin{matrix}{e_i^c}\\ e_i^\dag\end{matrix}\right), \quad
\nu_i=\left(\begin{matrix}\nu_i\\ {\nu_i}^\dag\end{matrix}\right)\quad i=1,2,3$

\underline{{\it charginos and neutralinos}}

$\tilde X^-_1=\left(\begin{matrix}\tilde \chi^-_1\\ \tilde {\chi}^{+\dag}_1\end{matrix}\right),\quad
\tilde X^+_1=\left(\begin{matrix}\tilde \chi^+_1\\ \tilde {\chi}^{-\dag}_1\end{matrix}\right), \quad
\tilde X^0_n=\left(\begin{matrix}\tilde \chi^0_n\\ \tilde {\chi}^{0\dag}_n\end{matrix}\right)$

\end{itemize}

\subsection{VEV's}

\begin{itemize}
\item {sneutrino VEV's \\
\quad $\left<\tilde \nu^c_{3}\right> \equiv \frac{1}{\sqrt 2} {v_R} \quad 
\epsilon_i=\frac{1}{2}Y_{\nu i3}v_R \quad \left<\tilde \nu_{i}\right> \equiv \frac{1}{\sqrt 2} {v_L}_i, \quad i=1,2,3$}

\item {Higgs VEV's \\
$\left< H_u^0\right> \equiv \frac{1}{\sqrt 2}v_u, \ \ \left< H_d^0\right> \equiv \frac{1}{\sqrt 2}v_d, \quad \tan \beta=v_u/v_d$}

\end{itemize}

\subsection{Relevant angles}

\begin{itemize}

\item{$\beta$ - Higgs VEVs ratio}
\begin{equation}
 \tan \beta=v_u/v_d
 \end{equation}

\item{$\theta_W$ - Weinberg angle}
\begin{equation}
\sin^2 \theta_W=0.22  \quad s_W=\sin \theta_W \quad c_W=\cos \theta_W
\end{equation}

\item{$\theta_R$ - $U_{B-L}$, $U_{3R}$ couplings ratio}

\begin{equation}
\cos \theta_R = \frac{g_R}{\sqrt{g_R^2+g_{BL}^2}} \ .
\end{equation}

\item{$\alpha$ - Higgs bosons rotation matrix}

\begin{equation}
\left(\begin{matrix}H_u^0\\H_d^0\end{matrix}\right)=
\left(\begin{matrix}v_u\\v_d\end{matrix}\right)+
\frac{1}{\sqrt{2}}R_{\alpha}\left(\begin{matrix}h^0\\H^0\end{matrix}\right)+
\frac{i}{\sqrt{2}}R_{\beta_0}\left(\begin{matrix}G^0\\\Gamma^0\end{matrix}\right) \ ,
\end{equation}

\begin{equation}
R_{\alpha}=\left( \begin{matrix}
\cos{\alpha}&\sin{\alpha}\\ -\sin{\alpha}&\cos{\alpha}
\end{matrix}\right),
\end{equation}

\item{$\phi_\pm$ - Chargino rotation matrix}

\begin{equation}
\tan 2\phi_-=2\sqrt{2}M_{W^\pm}\frac{\mu \cos \beta +M_2 \sin \beta}{\mu^2-M_2^2-2M_{W^\pm}^2
\cos 2\beta}
\label{bernard1}
\end{equation}
\begin{equation}
\tan 2\phi_+=2\sqrt{2}M_{W^\pm}\frac{\mu \sin \beta +M_2 \cos \beta}{\mu^2-M_2^2+2M_{W^\pm}^2
\cos 2\beta}
\label{bernard2}
\end{equation}

\item{Neutrino rotation matrix $V_{\text{PMNS}}$}

 The $3 \times 3$ Pontecorvo-Maki-Nakagawa-Sakata matrix is
\begin{eqnarray}
	V_\pmns &=& 
	\begin{pmatrix}
		c_{12} c_{13}
		&
		s_{12} c_{13}
		&
		s_{13} e^{-i \delta}
		\\
		-s_{12} c_{23} - c_{12} s_{23} s_{13} e^{i \delta}
		&
		c_{12} c_{23} - s_{12} s_{23} s_{13} e^{i \delta}
		&
		c_{13} s_{23}
		\\
		s_{12} s_{23} - c_{12}  c_{23} s_{13} e^{i \delta}
		&
		-c_{12} s_{23} - s_{12}  c_{23} s_{13} e^{i \delta}
		&
		c_{13} c_{23}
	\end{pmatrix}\nonumber\\ &&\times \text{diag}(1, e^{i \mathcal{A}/2}, 1) \ , \label{eq:32}
\end{eqnarray}
Values for the matrix terms can be found in \cite{Capozzi:2018ubv}.
\end{itemize}

\section{Neutralino decay rates}\label{appendix:B}

In \cite{Dumitru:2018jyb}, we computed the RPV decay rates of a general neutralino state $\tilde X_n^0$. the index $n$ indicates the neutralino species as follows:
\begin{equation}
{\tilde X}_1^0={\tilde X}_B^0,\quad  {\tilde X}_2^0={\tilde X}_W^0, \quad {\tilde X}_3^0={\tilde X}_{H_d}^0,
\quad {\tilde X}_4^0={\tilde X}_{H_u}^0, \quad {\tilde X}_5^0={\tilde X}_{\nu_{3a}}^0, \quad {\tilde X}_6^0={\tilde X}_{\nu_{3b}}^0.
\end{equation}
We reproduce the results here, for reference. 

\begin{enumerate}

\item{\boldmath{$\tilde X^0_n\rightarrow Z^0 \nu$}}

\begin{equation}\label{Neutralino_Decay_Rate1}
\Gamma_{{\tilde X}^0_n\rightarrow Z^0\nu_{i}}=
\frac{\Big(|{G_L}|_{{\tilde X}^0_n\rightarrow Z^0\nu_{i}}^2
+|{G_R}|_{{\tilde X}^0_n\rightarrow Z^0\nu_{i}}^2 \Big)
}{64\pi}
\frac{M_{{\tilde \chi}_n^0}^3}{M_{Z^0}^2}\left(1-\frac{M_{Z^0}^2}{M_{{\tilde \chi}_n^0}^2}\right)^2
\left(1+2\frac{M_{Z^0}^2}{M_{{\tilde \chi}^0_n}^2}\right),
\end{equation}
where
\begin{multline}
 {G_L}_{{\tilde X}^0_n\rightarrow Z^0 \nu_{i}}=
g_2\Big(\frac{1}{2c_W}\mathcal{N}_{n\>6+j}\mathcal{N}^*_{6+j\>6+i}-\frac{1}{c_W}\left(\frac{1}{2}+s_W^2\right)\mathcal{N}_{n\>4}\mathcal{N}^*_{6+i\>4} \Big)\\
+g_2\Big( \frac{1}{c_W}\left(\frac{1}{2}+s_W^2\right) \mathcal{N}^*_{n\>3}\mathcal{N}_{6+i\>3}\Big)
\end{multline}
and
\begin{multline}
{G_R}_{{\tilde X}^0_n\rightarrow Z^0 \nu_{i}}=g_2\Big( -\frac{1}{c_W}\left(\frac{1}{2}+s_W^2\right) \mathcal{N}_{n\>3}\mathcal{N}^*_{6+i\>3}\Big)\\
-{g_2}\Big[
\Big(-\frac{1}{2c_W}\mathcal{N}^*_{n\>6+j}\mathcal{N}_{6+j\>6+i}-\frac{1}{c_W}\left(\frac{1}{2}+s_W^2\right)\mathcal{N}^*_{n\>4}\mathcal{N}_{6+i\>4} \Big)
\end{multline}

\item{\boldmath{$\tilde X^0_n\rightarrow W^\mp \ell^\pm$}}

\begin{equation}\label{Neutralino_Decay_Rate2}
\Gamma_{{\tilde X}^0_n\rightarrow W^\mp \ell_i^\pm}=\frac{\Big(|{G_L}|_{{\tilde X}^0_n\rightarrow W^\pm \ell_i^\mp}^2+|{G_R}|_{{\tilde X}^0_n\rightarrow W^\pm \ell_i^\mp}^2\Big)}{64\pi}
\frac{M_{{\tilde \chi}_1^\pm}^3}{M_{W^\pm}^2}\left(1-\frac{M_{W^\pm}^2}{M_{{\tilde \chi}_n^0}^2}\right)^2
\left(1+2\frac{M_{W^\pm}^2}{M_{{\tilde \chi}_n^0}^2}\right),
\end{equation}
where
\begin{equation}
 {G_L}_{{\tilde X}^0_n\rightarrow W^- \ell_i^+}=-{G_R}_{{\tilde X}^0_n\rightarrow W^+ \ell_i^-}=\frac{g_2}{\sqrt{2}}\Big[\mathcal{N}_{n\>4}\mathcal{V}^*_{2+i\>2}+\sqrt{2}\mathcal{V}^*_{2+i\>1}\mathcal{N}_{n\>2}\Big]
\end{equation}
and
\begin{equation}
{G_R}_{{\tilde X}^0_n\rightarrow W^- \ell_i^+}=-{G_L}_{{\tilde X}^0_n\rightarrow W^+ \ell_i^-}=\frac{g_2}{\sqrt{2}}\Big[-\mathcal{U}_{2+i\>2+j}\mathcal{N}^*_{n\>6+j}-\mathcal{U}_{2+i\>2}\mathcal{N}^*_{n\>3}+\sqrt{2}\mathcal{N}^*_{n\>2}\mathcal{U}_{2+i\>1}\Big]
\end{equation}

\item{\boldmath{$\tilde X^0_n\rightarrow h^0 \nu$}}
\begin{equation}\label{Neutralino_Decay_Rate3}
\Gamma_{{\tilde X}^0_n\rightarrow h^0\nu_{i}}=\frac{\Big(|{G_L}|_{{\tilde X}^0_n\rightarrow h^0\nu_{i}}^2+|{G_R}|_{{\tilde X}^0_n\rightarrow h^0\nu_{i}}^2\Big)}{64\pi}
M_{{\tilde \chi}_n^0}\left(1-\frac{M_{h^0}^2}{M_{{\tilde \chi}_n^0}^2}\right)^2
\end{equation}
where
\begin{multline}
{G_L}_{{\tilde X}^0_n\rightarrow h^0 \nu_{i} }=\frac{g_2}{{2}}\Big(
\cos \alpha (\mathcal{N}^*_{n\>4}\mathcal{N}^*_{6+i\>2}+\mathcal{N}^*_{6+i\>4}\mathcal{N}_{n\>2}^*)+\sin \alpha (\mathcal{N}^*_{n\>3}\mathcal{N}^*_{6+i\>2}+\mathcal{N}^*_{6+i\>3}\mathcal{N}_{n\>2}^*)\Big)\\
-\frac{g'}{{2}}\Big(
\cos\alpha \left(\sin \theta_R(\mathcal{N}^*_{n\>4}\mathcal{N}^*_{6+i\>1}+\mathcal{N}^*_{6+i\>4}\mathcal{N}^*_{n\>1})+\cos \theta_R(\mathcal{N}^*_{n\>4}\mathcal{N}^*_{6+i\>5}+\mathcal{N}^*_{6+i\>4}\mathcal{N}^*_{n\>5})   \right)\\
+\sin \alpha \left( \sin \theta_R(\mathcal{N}^*_{n\>3}\mathcal{N}^*_{6+i\>1}+\mathcal{N}^*_{6+i\>3}\mathcal{N}^*_{n\>1})+\cos \theta_R(\mathcal{N}^*_{n\>3}\mathcal{N}^*_{6+i\>5}+\mathcal{N}^*_{6+i\>3}\mathcal{N}^*_{n\>5}) \right)\Big)\\
+\frac{1}{\sqrt 2}Y_{\nu i3}\cos\alpha \Big(\mathcal{N}^*_{n\>6+j}\mathcal{N}^*_{6+i\>6}
+\mathcal{N}^*_{6+i\>6+j}
\mathcal{N}^*_{n\>6}\Big)
\end{multline}
and
\begin{multline}
{G_R}_{{\tilde X}^0_n\rightarrow h^0 \nu_{i} }=\frac{g_2}{{2}}\Big(
\cos \alpha (\mathcal{N}_{n\>4}\mathcal{N}_{6+i\>2}+\mathcal{N}_{6+i\>4}\mathcal{N}_{n\>2})+\sin \alpha (\mathcal{N}_{n\>3}\mathcal{N}_{6+i\>2}+\mathcal{N}_{6+i\>3}\mathcal{N}_{n\>2})\Big)\\
+\frac{g'}{{2}}\Big(\cos\alpha \left(\sin \theta_R(\mathcal{N}_{n\>4}\mathcal{N}_{1\>6+i}+\mathcal{N}_{6+i\>4}\mathcal{N}_{n\>1})+\cos \theta_R(\mathcal{N}_{n\>4}\mathcal{N}_{6+i\>5}+\mathcal{N}_{6+i\>4}\mathcal{N}_{n\>5})   \right)\\
+\sin \alpha \left( \sin \theta_R(\mathcal{N}_{n\>3}\mathcal{N}_{6+i\>1}+\mathcal{N}_{6+i\>3}\mathcal{N}_{n\>1})+\cos \theta_R(\mathcal{N}_{n\>3}\mathcal{N}_{6+i\>5}+\mathcal{N}_{6+i\>3}\mathcal{N}_{n\>5}) \right)\Big)\\
+\Big(\mathcal{N}_{n\>6+j}\mathcal{N}_{6+i\>6}
+\frac{1}{\sqrt 2}Y_{\nu i3}\cos\alpha \Big(\mathcal{N}_{6+i\>6+j}
\mathcal{N}_{n\>6}\Big)
\end{multline}

\end{enumerate}
The matrices $\mathcal{U}$, $\mathcal{V}$ and $\mathcal{N}$ matrices rotate the gaugino eigenstates into the neutralino and chargino mass eigenstates. They are presented in Appendices B.1 and B.2 of \cite {Dumitru:2018jyb}. 

Note that in all cases in Appendix A and B above, we sum over $j=1,2,3$.

\end{appendices}

\end{document}